\documentclass[journal]{IEEEtran}

\IEEEoverridecommandlockouts

\usepackage{amsmath,amsthm,relsize}
\usepackage{amsfonts, amssymb, cuted}
\usepackage{cleveref}
\usepackage{mathtools}
\usepackage[boxed]{algorithm}
\usepackage[noend]{algpseudocode}
\usepackage{cite}
\usepackage{xcolor}
\usepackage{soul}
\usepackage{graphicx}
\usepackage{tikz}
\usepackage{pgfplotstable}
\usepackage{pgfplots}
\usepackage{nicefrac}
\usepackage{enumitem}
\usepackage{support-caption}
\usepackage{subcaption}
\usepackage[font=footnotesize]{caption}
\usepackage{multirow}
\pgfplotsset{compat=1.15}
\usepackage{relsize}
\usetikzlibrary{patterns}
\usepackage{acro}
\usepackage{todonotes}
\usepackage{color,soul}
\usepackage{bbm}
\usepackage{multirow}
\usepackage{makecell}

\newcommand{\vbar}{\raisebox{.17ex}{\rule{.04em}{1.35ex}}}
\newcommand{\vbarind}{\raisebox{.01ex}{\rule{.04em}{1.1ex}}}
\newcommand{\R}{\ifmmode{\rm I}\hspace{-.2em}{\rm R} \else ${\rm I}\hspace{-.2em}{\rm R}$ \fi}
\newcommand{\T}{\ifmmode{\rm I}\hspace{-.2em}{\rm T} \else ${\rm I}\hspace{-.2em}{\rm T}$ \fi}
\newcommand{\N}{\ifmmode{\rm I}\hspace{-.2em}{\rm N} \else \mbox{${\rm I}\hspace{-.2em}{\rm N}$} \fi}
\newcommand{\B}{\ifmmode{\rm I}\hspace{-.2em}{\rm B} \else \mbox{${\rm I}\hspace{-.2em}{\rm B}$} \fi}
\newcommand{\Hil}{\ifmmode{\rm I}\hspace{-.2em}{\rm H} \else \mbox{${\rm I}\hspace{-.2em}{\rm H}$} \fi}
\newcommand{\C}{\ifmmode\hspace{.2em}\vbar\hspace{-.31em}{\rm C} \else \mbox{$\hspace{.2em}\vbar\hspace{-.31em}{\rm C}$} \fi}
\newcommand{\Cind}{\ifmmode\hspace{.2em}\vbarind\hspace{-.25em}{\rm C} \else \mbox{$\hspace{.2em}\vbarind\hspace{-.25em}{\rm C}$} \fi}
\newcommand{\Q}{\ifmmode\hspace{.2em}\vbar\hspace{-.31em}{\rm Q} \else \mbox{$\hspace{.2em}\vbar\hspace{-.31em}{\rm Q}$} \fi}
\newcommand{\Z}{\ifmmode{\rm Z}\hspace{-.28em}{\rm Z} \else ${\rm Z}\hspace{-.28em}{\rm Z}$ \fi}

\DeclareAcronym{AWGN}{
    short = AWGN,
    long = additive white Gaussian noise,
    list = Additive White Gaussian Noise,
    tag = abbrev
}

\DeclareAcronym{ADMM}{
    short = ADMM,
    long = alternating direction method of multipliers,
    list = Alternating Direction Method of Multipliers,
    tag = abbrev
}

\DeclareAcronym{MGMC}{
    short = MGMC,
    long = multi-group multi-casting,
    list = multi-group multi-casting,
    tag = abbrev
}

\DeclareAcronym{SGMC}{
    short = SGMC,
    long = single-group multi-casting,
    list = single-group multi-casting,
    tag = abbrev
}
\DeclareAcronym{AoA}{
    short = AoA,
    long = angle-of-arrival,
    list = Angle-of-Arrival,
    tag = abbrev
}

\DeclareAcronym{AoD}{
    short = AoD,
    long = angle-of-departure,
    list = Angle-of-Departure,
    tag = abbrev
}

\DeclareAcronym{KKT}{
    short = KKT,
    long = Karush-Kuhn-Tucker,
    list = Karush-Kuhn-Tucker,
    tag = abbrev
}

\DeclareAcronym{MMF}{
    short = MMF,
    long = max-min-fairness,
    list = max-min-fairness,
    tag = abbrev
}

\DeclareAcronym{WMMF}{
    short = WMMF,
    long = weighted max-min-fairness,
    list = max-min-fairness,
    tag = abbrev
}

\DeclareAcronym{BB}{
    short = BB,
    long = base band,
    list = Base Band,
    tag = abbrev
}

\DeclareAcronym{BC}{
    short = BC,
    long = broadcast channel,
    list = Broadcast Channel,
    tag = abbrev
}

\DeclareAcronym{BS}{
    short = BS,
    long = base station,
    list = Base Station,
    tag = abbrev
}

\DeclareAcronym{BR}{
    short = BR,
    long = best response,
    list = Best Response, 
    tag = abbrev
}

\DeclareAcronym{CB}{
    short = CB,
    long = coordinated beamforming,
    list = Coordinated Beamforming,
    tag = abbrev
}

\DeclareAcronym{CC}{
    short = CC,
    long = coded caching,
    list = Coded Caching,
    tag = abbrev
}

\DeclareAcronym{CE}{
    short = CE,
    long = channel estimation,
    list = Channel Estimation,
    tag = abbrev
}

\DeclareAcronym{CoMP}{
    short = CoMP,
    long = coordinated multi-point transmission,
    list = Coordinated Multi-Point Transmission,
    tag = abbrev
}

\DeclareAcronym{CRAN}{
    short = C-RAN,
    long = cloud radio access network,
    list = Cloud Radio Access Network,
    tag = abbrev
}

\DeclareAcronym{CSE}{
    short = CSE,
    long = channel specific estimation,
    list = Channel Specific Estimation,
    tag = abbrev
}

\DeclareAcronym{CSI}{
    short = CSI,
    long = channel state information,
    list = Channel State Information,
    tag = abbrev
}

\DeclareAcronym{CSIT}{
    short = CSIT,
    long = channel state information at the transmitter,
    list = Channel State Information at the Transmitter,
    tag = abbrev
}

\DeclareAcronym{CU}{
    short = CU,
    long = central unit,
    list = Central Unit,
    tag = abbrev
}

\DeclareAcronym{D2D}{
    short = D2D,
    long = device-to-device,
    list = Device-to-Device,
    tag = abbrev
}

\DeclareAcronym{DE-ADMM}{
    short = DE-ADMM,
    long = direct estimation with alternating direction method of multipliers,
    list = Direct Estimation with Alternating Direction Method of Multipliers,
    tag = abbrev
}

\DeclareAcronym{DE-BR}{
    short = DE-BR,
    long = direct estimation with best response,
    list = Direct Estimation with Best Response,
    tag = abbrev
}

\DeclareAcronym{DE-SG}{
    short = DE-SG,
    long = direct estimation with stochastic gradient,
    list = Direct Estimation with Stochastic Gradient,
    tag = abbrev
}

\DeclareAcronym{DFT}{
	short = DFT,
	long = discrete fourier transform,
	list = Discrete Fourier Transform,
	tag = abbrev
}

\DeclareAcronym{DoF}{
    short = DoF,
    long = degrees of freedom,
    list = Degrees of Freedom,
    tag = abbrev
}

\DeclareAcronym{DL}{
    short = DL,
    long = downlink,
    list = Downlink,
    tag = abbrev
}

\DeclareAcronym{GD}{
	short = GD, 
	long = gradient descent,
	list = Gradeitn Descent,
	tag = abbrev
}

\DeclareAcronym{IBC}{
    short = IBC,
    long = interfering broadcast channel,
    list = Interfering Broadcast Channel,
    tag = abbrev
}

\DeclareAcronym{i.i.d.}{
    short = i.i.d.,
    long = independent and identically distributed,
    list = Independent and Identically Distributed,
    tag = abbrev
}

\DeclareAcronym{JP}{
    short = JP,
    long = joint processing,
    list = Joint Processing,
    tag = abbrev
}

\DeclareAcronym{LOS}{
	short = LOS,
	long = line-of-sight,
	list = Line-of-Sight,
	tag = abbrev
}

\DeclareAcronym{LS}{
    short = LS,
    long = least squares,
    list = Least Squares,
    tag = abbrev
}

\DeclareAcronym{LTE}{
    short = LTE,
    long = Long Term Evolution,
    tag = abbrev
}

\DeclareAcronym{LTE-A}{
    short = LTE-A,
    long = Long Term Evolution Advanced,
    tag = abbrev
}

\DeclareAcronym{MIMO}{
    short = MIMO,
    long = multiple-input multiple-output,
    list = Multiple-Input Multiple-Output,
    tag = abbrev
}

\DeclareAcronym{MISO}{
    short = MISO,
    long = multiple-input single-output,
    list = Multiple-Input Single-Output,
    tag = abbrev
}

\DeclareAcronym{MAC}{
    short = MAC,
    long = multiple access channel,
    list = Multiple Access Channel,
    tag = abbrev
}

\DeclareAcronym{MSE}{
    short = MSE,
    long = mean-squared error,
    list = Mean-Squared Error,
    tag = abbrev
}

\DeclareAcronym{MMSE}{
    short = MMSE,
    long = minimum mean-squared error,
    list = Minimum Mean-Squared Error,
    tag = abbrev
}

\DeclareAcronym{mmWave}{
	short = mmWave,
	long = millimeter wave,
	list = Millimeter Wave,
	tag = abbrev
}

\DeclareAcronym{MU-MIMO}{
    short = MU-MIMO,
    long = multi-user \ac{MIMO},
    list = Multi-User \ac{MIMO},
    tag = abbrev
}

\DeclareAcronym{OTA}{
    short = OTA,
    long = over-the-air,
    list = Over-the-Air,
    tag = abbrev
}

\DeclareAcronym{PSD}{
    short = PSD,
    long = positive semidefinite,
    list = Positive Semidefinite,
    tag = abbrev
}

\DeclareAcronym{QoS}{
	short = QoS,
	long = quality of service,
	list = Quality of Service,
	tag = abbrev
}

\DeclareAcronym{RCP}{
	short = RCP,
	long = remote central processor,
	list = Remote Central Processor,
	tag = abbrev
}

\DeclareAcronym{RRH}{
    short = RRH,
    long = remote radio head,
    list = Remote Radio Head,
    tag = abbrev
}

\DeclareAcronym{RSSI}{
    short = RSSI,
    long = received signal strength indicator,
    list = Received Signal Strength Indicator,
    tag = abbrev
}

\DeclareAcronym{RX}{
	short = RX,
	long = receiver,
	list = Receiver,
	tag = abbrev
}

\DeclareAcronym{SCA}{
    short = SCA,
    long = successive-convex-approximation,
    list = Successive-Convex-Approximation,
    tag = abbrev
}

\DeclareAcronym{SG}{
    short = SG,
    long = stochastic gradient,
    list = Stochastic Gradient,
    tag = abbrev
}

\DeclareAcronym{SIC}{
    short = SIC,
    long = successive interference cancellation,
    list = Successive Interference Cancellation,
    tag = abbrev
}

\DeclareAcronym{SNR}{
    short = SNR,
    long = signal-to-noise-ratio,
    list = Signal-to-Noise Ratio,
    tag = abbrev
}

\DeclareAcronym{SDR}{
    short = SDR,
    long = semi-definite-relaxation,
    list = semi-definite-relaxation,
    tag = abbrev
}

\DeclareAcronym{SINR}{
    short = SINR,
    long = signal-to-interference-plus-noise ratio,
    list = Signal-to-Interference-plus-Noise Ratio,
    tag = abbrev
}

\DeclareAcronym{SOCP}{
	short = SOCP, 
	long = second order cone program,
	list = Second Order Cone Program,
	tag = abbrev
}

\DeclareAcronym{SSE}{
    short = SSE,
    long = stream specific estimation,
    list = Stream Specific Estimation,
    tag = abbrev
}

\DeclareAcronym{SVD}{
	short = SVD,
	long = singular value decomposition,
	list = Singular Value Decomposition,
	tag = abbrev
}

\DeclareAcronym{TDD}{
	short = TDD,
	long = time division duplex,
	list = Time Division Duplex,
	tag = abbrev
}

\DeclareAcronym{TX}{
	short = TX,
	long = transmitter,
	list = Transmitter,
	tag = abbrev
}

\DeclareAcronym{UE}{
    short = UE,
    long = user equipment,
    list = User Equipment,
    tag = abbrev
}

\DeclareAcronym{UL}{
    short = UL,
    long = uplink,
    list = Uplink,
    tag = abbrev
}

\DeclareAcronym{ULA}{
	short = ULA,
	long = uniform linear array,
	list = Uniform Linear Array,
	tag = abbrev
}

\DeclareAcronym{UPA}{
    short = UPA,
    long = uniform planar array,
    list = Uniform Planar Array,
    tag = abbrev
}

\DeclareAcronym{WMMSE}{
    short = WMMSE,
    long = weighted minimum mean-squared error,
    list = Weighted Minimum Mean-Squared Error,
    tag = abbrev
}

\DeclareAcronym{WMSEMin}{
    short = WMSEMin,
    long = weighted sum \ac{MSE} minimization,
    list = Weighted sum \ac{MSE} Minimization,
    tag = abbrev
}

\DeclareAcronym{WBAN}{
	short = WBAN,
	long = wireless body area network,
	list = Wireless Body Area Network,
	tag = abbrev
}

\DeclareAcronym{WSRMax}{
    short = WSRMax,
    long = weighted sum rate maximization,
    list = Weighted Sum Rate Maximization,
    tag = abbrev
}

\newtheorem{exmp}{Example}%[section]
\theoremstyle{definition}

\newtheorem{rem}{Remark}

\newcommand{\CA}[0]{{\mathcal{A}}}
\newcommand{\CB}[0]{{\mathcal{B}}}
\newcommand{\CC}[0]{{\mathcal{C}}}

\newcommand{\CH}[0]{{\mathcal{H}}}
\newcommand{\CI}[0]{{\mathcal{I}}}

\newcommand{\CK}[0]{{\mathcal{K}}}
\newcommand{\CL}[0]{{\mathcal{L}}}

\newcommand{\CS}[0]{{\mathcal{S}}}
\newcommand{\CT}[0]{{\mathcal{T}}}
\newcommand{\CU}[0]{{\mathcal{U}}}
\newcommand{\CV}[0]{{\mathcal{V}}}

\newcommand{\Bm}[0]{{\mathbf{m}}}

\newcommand{\Bp}[0]{{\mathbf{p}}}

\newcommand{\Br}[0]{{\mathbf{r}}}

\newcommand{\Bu}[0]{{\mathbf{u}}}
\newcommand{\Bv}[0]{{\mathbf{v}}}

\newcommand{\BA}[0]{{\mathbf{A}}}

\newcommand{\BQ}[0]{{\mathbf{Q}}}
\newcommand{\BR}[0]{{\mathbf{R}}}

\newcommand{\Sfc}[0]{{\mathsf{c}}}

\newcommand{\Sff}[0]{{\mathsf{f}}}

\newcommand{\Sfl}[0]{{\mathsf{l}}}

\newcommand{\Sfn}[0]{{\mathsf{n}}}

\newcommand{\Sfx}[0]{{\mathsf{x}}}

\newcommand{\SfL}[0]{{\mathsf{L}}}

\usepackage{titlesec}
\titlespacing\section{3pt}{6pt plus 4pt minus 2pt}{6pt plus 2pt minus 2pt}
\titlespacing\subsection{3pt}{4pt plus 4pt minus 2pt}{4pt plus 2pt minus 2pt}
\titlespacing\subsubsection{3pt}{3pt plus 4pt minus 2pt}{0pt plus 2pt minus 3pt}

\title{Fairness Scheduling for Coded Caching in Multi-AP Wireless Local Area Networks}

\begin{document}

\author{\IEEEauthorblockN{Kagan Akcay,
%\IEEEauthorrefmark{1}, 
MohammadJavad Salehi, %\IEEEauthorrefmark{2}, 
and Giuseppe Caire}%\IEEEauthorrefmark{1}} 
\\
%\IEEEauthorblockA{
%    \IEEEauthorrefmark{1} Electrical Engineering and Computer Science Department, Technische Universit\"at Berlin, 10587 Berlin, Germany\\
%    \IEEEauthorrefmark{2} Centre for Wireless Communications, University of Oulu, 90570 Oulu, Finland \\
%    \textrm{kagan.akcay@tu-berlin.de \quad mohammadjavad.salehi@oulu.fi \quad caire@tu-berlin.de}
%    }
\thanks{
K. Akcay and G. Caire are with the Electrical Engineering and Computer Science Department, Technische Universit\"at Berlin, 10587 Berlin, Germany. E-mails: kagan.akcay@tu-berlin.de and caire@tu-berlin.de. %Their work is partially funded  
%BMBF Germany in the program of ``Souverän. Digital. Vernetzt.'' Joint Project 6G-RIC (Project IDs 16KISK030)
MJ Salehi is with the Centre for Wireless Communications, University of Oulu, 90570 Oulu, Finland. E-mail:  mohammadjavad.salehi@oulu.fi. %His work is supported by the Academy of Finland, under Grants No. 346208 (6G Flagship program) and 343586 (CAMAIDE), and by the Nokia Foundation through the Jorma-Ollila grant. 
Part of this work has been presented at the IEEE GLOBECOM 2023, in Kuala Lumpur, Malaysia~\cite{akcay_gc}.}
}

\maketitle

\begin{abstract}
Coded caching (CC) schemes exploit the cumulative cache memory in the user's devices and 
linear coding to turn unicast traffic (individual file requests) into a multicast transmission.
%For the original setting proposed by Maddah-Ali and Niesen, as well as
%comprising $K$ users receiving information 
%from a server via a shared multicast link, CC yields an $O(K)$ rate gain with respect to conventional uncoded caching with the same per-user memory. While several information-theoretic optimality results 
For a variety of carefully crafted network topologies, information theoretic optimality results showing large gains with respect to conventional uncoded caching have been proved. However, the benefits and suitability of CC for practical scenarios, such as content streaming over existing wireless networks, have not yet been fully demonstrated. In this work, we study CC for on-demand video streaming over large Wireless Local Area Networks, where multiple users are served simultaneously by multiple spatially distributed access points (APs). Users sequentially request video ``chunks" of their selected video file in a given content library. To enhance practical applicability, we propose a CC solution with a completely asynchronous, decentralized, and location-independent cache placement, paired with an ``over IP'' delivery mechanism operating in higher network layers and leaving the underlying physical and MAC layers untouched. For this CC scheme, 
we formulate the region of {\em achievable goodput}, defined as the number of video chunks per unit time
delivered in the users' playback buffer, and formulate the goodput fairness problem in terms of a convex maximization problem. %\color{red} 
We %also consider 
then propose an associated dynamic scheduling algorithm that provably achieves the optimal fairness point in stationary conditions with reduced complexity, % \color{blue}Then, we consider reduced complexity scheduling strategies that are scalable to large networks.\color{black} 
and then introduce a heuristic to further reduce the complexity, ultimately achieving a favorable trade-off between performance and complexity.
%\color{black} (( Last sentence is almost the same as in the intro, but I think it should be fine. Because here, saying something like 'scalable to a large network' or 'scalable to much larger networks' doesn't seem fine, I think. There are two complexity reductions, and it seems tricky. ))
We illustrate the performance of the proposed solutions by comparing them with standard baseline schemes such as 1) conventional (prefix) caching; 2) collision avoidance by allocating APs on orthogonal sub-channels with a certain spatial reuse; and 3) a CSMA-inspired solution where the APs trigger their activity via exponentially distributed waiting times and obey the ready-to-send/clear-to-send distributed coordination function. Our results show that CC can be implemented as an over IP solution in a scalable, compatible way with existing WLAN standards, achieving significant gains over all the considered baseline approaches. 
\end{abstract}

\begin{IEEEkeywords}
coded caching, multi-AP WLANs, 
video streaming, 
fairness scheduling.
\end{IEEEkeywords}

\section{Introduction}

\subsection{Motivation and Prior Art}

The increasing amount of data traffic, especially driven by multimedia applications, necessitates the development of new communication techniques~\cite{rajatheva2020white}. One interesting resource is on-device memory; it is cheap and can be used to proactively store a large part of multimedia content, e.g., for video-on-demand and extended reality applications~\cite{salehi2022enhancing}. As a result, many researchers have considered the efficient use of onboard memory in the context of \emph{caching}. Pioneering works in this regard introduced femtocaching~\cite{shanmugam2013femtocaching} and F-RAN models~\cite{park2016joint}. A major theoretical breakthrough is the introduction of coded caching (CC)~\cite{maddah2014fundamental}, which shows a gain factor in the content delivery rate proportional to the {\em cumulative} cache size in the network by leveraging the differences in user cache memories to turn unicast data traffic (i.e., satisfying individual user file demands) into multicast transmissions.
%CC leverages a server's ability to multicast common (coded) messages to a large number of users, cleverly designed combinatorial network codes, and the information stored in user caches to turn unicast data traffic (i.e., satisfying individual user file demands)into multicast transmissions. 

Since its introduction, CC has generated a vast and influential body of work at the intersection of information theory, network coding, and algebraic/combinatorial coding theory. For example, 
% and combinatorial design. %, including several best-paper award–winning contributions. The relevance and intellectual importance of this line of research are therefore beyond dispute. However, despite this success, the impact of CC on real-world content delivery systems has so far remained limited.
numerous works have built on the original CC framework of~\cite{maddah2014fundamental}, e.g., for wireless~\cite{tolli2017multi,NaseriTehrani2024Cache-AidedOptimization}, multi-server~\cite{shariatpanahi2016multi}, multi-antenna~\cite{shariatpanahi2018physical,salehi2021MIMO}, device-to-device (D2D)~\cite{ji2015fundamental}, shared-cache~\cite{parrinello2019fundamental,parrinello2020extending}, multi-access~\cite{serbetci2019multi}, dynamic~\cite{abolpour2023cache,Abolpour2024ResourceBehavior}, and combinatorial~\cite{brunero2022fundamental} networks.

While these works are mathematically elegant and often yield information-theoretic optimality results~\cite{kai_optimal,caching_tightbound,lampiris2018resolving},
their application to practical wireless systems remains challenging. In particular, most theoretical works on coded caching applied to wireless systems involve modifications to the physical layer that are incompatible with current wireless standards, such as the ability to perform CC precoding directly in the complex signal domain (e.g., multi-user MIMO precoding)~\cite{cache-interference,lampiris2018adding,salehi2020lowcomplexity}. Such modifications are incompatible with the structure of contemporary wireless systems, where CC is implemented at the ``application layer’’ (i.e., above the IP layer), while the PHY and MAC layers are standardized independently (e.g., IEEE 802.11~\cite{khorov2018tutorial} and 3GPP~\cite{ghosh20195g}), and must support %a wider range of applications and traffic types.
a wide variety of services beyond content delivery.
%most CC formulations for wireless networks assume that the physical layer (e.g., multi-user MIMO precoding) can be co-designed with the CC scheme~\cite{cache-interference,lampiris2018adding,salehi2020lowcomplexity}. Such assumptions are incompatible with the structure of contemporary wireless systems, where CC is implemented at the ``application layer’’, while the PHY and MAC layers are standardized independently (e.g., IEEE 802.11~\cite{khorov2018tutorial} and 3GPP~\cite{ghosh20195g}) and must support a wide variety of services beyond content delivery.
As a result, there exists a fundamental gap between the theoretical development of CC in wireless networks and the practice of on-demand %media 
video streaming, which today operates almost entirely above the IP layer and relies only on limited adaptations of scheduling and queue management mechanisms.

Recognizing this practical constraint,~\cite{mozhgan} considered CC ``over IP'', i.e., above an existing network layer capable of multicast routing, by considering 
a network model based on carrier-sense multiple access (CSMA), 
reminiscent of a WiFi Wireless Local Area Network (WLAN) 
with multiple access points (APs). 
In~\cite{mozhgan},  a server transmits data to multiple spatially distributed users through multiple APs where the signal transmitted by each AP is received by all the users within a certain radius. 
A collision-type interference model consistent with CSMA is used, such that a packet is lost if a user 
receives the superposition of concurrent packets from different APs above a certain interference threshold (a similar model is elaborated in this paper, see Section~\ref{section:sys_model}). 
Using the multiround delivery method of~\cite{caire} (which was later proved by~\cite{parrinello2019fundamental} to be optimal for the shared-cache model), \cite{mozhgan} considered a graph coloring approach to construct a spatial reuse pattern for the APs, with optimized AP-user assignment to minimize delivery time. Then, a heuristic scheme dubbed  ``Avalanche'' was introduced to leverage the fact that, as APs finish serving their assigned users, they stop transmitting and they may free other users from interference. As a consequence, AP activation propagates through the network in a sort of avalanche effect, constrained by the standard WiFi distributed clear-to-send/ready-to-send coordination function \cite{bianchi2000performance}.

While~\cite{mozhgan} established the relevance of CC over IP for WLAN-type networks and demonstrated promising performance gains, its solutions are inherently heuristic and do not optimize a well-defined network-wide performance metric.
Moreover, \cite{mozhgan} as well as the great majority of information-theoretic papers on CC (e.g., see \cite{maddah2014fundamental,tolli2017multi,shariatpanahi2016multi,shariatpanahi2018physical,salehi2021MIMO,ji2015fundamental,parrinello2019fundamental,parrinello2020extending,serbetci2019multi,abolpour2023cache,brunero2022fundamental,kai_optimal,caching_tightbound,lampiris2018resolving,salehi2020lowcomplexity,lampiris2018adding}, as well as many others) focus on minimizing the worst-case (over the user demands) 
delivery time for the whole set of requested files. 
%In these works, the {\em rate} of the CC scheme is defined as the length of the coded transmission 
%necessary to deliver to each user its requested content data unit (i.e., a  file in the library), normalized by the length of one data unit.
%In this classical version of the CC problem, it does not matter if a user gets their requested file 
%at the beginning or at the very end of the delivery time, as long as the delivery time over the {\em whole network}, under the worst-case demand configuration, is minimized. 
%While this problem formulation may be suited to {\em file download} applications, 
%it is  certainly not suited to {\em media streaming}. 
%In streaming (e.g., video), a very long content file (a video with a duration of several minutes) is chopped into equal sized ``chunks'', which are fetched sequentially via HTTP requests from the video client running on the user device \cite{Http_videochunks}. 
%The chunks contain a relatively small segment of the video, of a duration of a few seconds, 
%and they must be delivered {\em sequentially} at a rate (slightly) higher than the playback rate at which 
%the client buffer is emptied by the video player application~\cite{bethanabhotla2014adaptive,caire_videobuffer}.  
%Hence, the most relevant performance metric is the delivery rate, expressed as the long-term time-averaged number of chunks sequentially delivered per unit of time. 
%
In this classical version of the CC problem, it does not matter whether a user receives its requested file 
at the beginning or at the end of the delivery time, as long as the delivery time over the {\em whole network}, under the worst-case demand configuration, is minimized. 
While this formulation may be suited to {\em file download} applications, 
it is certainly not suited to {\em media streaming}. 
In streaming (e.g., video), a very long content file (a video with a duration of several minutes) is divided into equal-sized ``chunks'', which are fetched sequentially via HTTP requests from the video client running on the user device~\cite{Http_videochunks}. 
The chunks contain relatively short segments of the video and must be delivered {\em sequentially} at a rate (slightly) higher than the playback rate at which 
the client buffer is emptied by the video player application~\cite{bethanabhotla2014adaptive,caire_videobuffer}.  
Hence, the most relevant performance metric is the delivery rate, expressed as the long-term time-averaged number of chunks delivered per unit time.

Furthermore, since multiple users share the same wireless resources and may not simultaneously experience high rates, it is essential to ensure fairness across users and to provide fair rates, a principle that underlies virtually all modern wireless scheduling algorithms~\cite{tse-fairness,georgiadis2006resource,queue2,queue_exact}.

%Furthermore, since multiple users share the same wireless resources, it is essential not only to maximize efficiency but also to ensure fairness across users, a principle that underlies virtually all modern wireless scheduling algorithms, from proportional fairness to max-min fairness~\cite{tse2001smart,georgiadis2006resource,queue2,queue_exact}.

\subsection{Our Contributions}

%{\color{red}
%Motivated by the above observations, this paper addresses the problem of CC-based media streaming over multi-AP WiFi-type networks from a fundamentally different toggle perspective than existing works.
%}
%We focus on the {\em fairness scheduling problem} in a multi-AP network obeying a collision-type {\em WiFi-like} interference 
%model.\footnote{The interference collision model is a convenient idealization of the actual WiFi MAC layer, and it is chosen here for analytical tractability.} 
%In CC, video chunks are partitioned into subpackets placed in user caches during the placement phase, and a chunk becomes useful to the video player only after all its subpackets have been delivered. 
%{\color{red}
%We define the {\em goodput} of a user as the long-term time-averaged rate at which complete chunks are delivered to its playback %buffer, and adopt it as the central performance metric of this work.
%}

Motivated by the above observations, this paper addresses the problem of CC-based video streaming over multi-AP WiFi-type networks from a fundamentally different perspective than existing works.
%This work focuses 
We focus on the {\em fairness scheduling problem} in a multi-AP network obeying a collision-type {\em WiFi-like} interference 
model.\footnote{The interference collision model is a convenient idealization of the actual WiFi MAC layer, and it is chosen here for the sake of analytical
tractability.} In CC, some segments of the chunks, referred to as {\em subpackets}, 
are placed in the user cache memory during the cache placement phase. However, 
a chunk is useless to the video player until {\em all} the subpackets forming the chunk are delivered. 
In the following, we refer to the (time-averaged) rate at which chunks are made available to the user as the {\em goodput} of the user.

We obtain the achievable users' goodput region as the convex hull of all {\em instantaneous} delivery rate vectors. The fairness scheduling problem is then formulated as the maximization of a concave, component-wise non-decreasing network utility function~\cite{fairness} over the goodput region. To enable continuous streaming, user goodputs must exceed the video playback rate~\cite{bethanabhotla2014adaptive,caire_videobuffer}. The utility function that directly enforces this requirement is 
hard fairness, which maximizes the minimum user goodput. However, when proportional fairness is employed, at the expense of a reduction in the goodputs of some users, most users can stream at significantly higher quality when adaptive quality coding is used~\cite{lefteris-adaptivequality}. 
Accordingly, in this paper we consider both hard fairness and proportional fairness as network utility functions.

Although the formulated fairness scheduling problem is convex, its direct solution is not applicable to dynamic networks, and the number of variables (i.e., the number of maximal instantaneous rate vectors) grows faster than exponentially with the network size. To enable applicability to dynamic networks, we adopt the Lyapunov Drift-Plus-Penalty (DPP) method~\cite{georgiadis2006resource,queue2,queue_exact} and exploit symmetry among users with identical cache profiles to significantly reduce computational complexity. 
For larger networks, where exact optimality becomes intractable, we introduce a principled heuristic that further reduces complexity by prioritizing transmissions to users with large queue backlogs within the DPP framework, achieving a favorable trade-off between performance and complexity.

We demonstrate the effectiveness of the proposed framework through extensive simulations. As baseline schemes for comparison, we consider: 
1) conventional (i.e., prefix) caching; 
2) a spatial reuse–based scheme, where APs are allocated orthogonal transmission resources (e.g., orthogonal WiFi channels~\cite{zubow2016bigap,bhartia2019clientmarshal}) to avoid mutual interference, and where each AP maximizes its weighted sum rate according to the DPP method~\cite{georgiadis2006resource,queue2,queue_exact}; 
and 3) a CSMA-inspired solution in which APs are activated according to i.i.d. exponentially distributed waiting times and transmit only if they do not create interference with users already being served by other active APs, following the standard distributed coordination function.

Our results show that: 
1) for all schemes, CC yields a significant gain compared to conventional (e.g., prefix) caching, due to its ability to exploit multicast transmissions; and 
2) the proposed optimal scheme (when computationally feasible --e.g., with less than ten APs in the network) and the proposed principled heuristic (for very large networks) significantly outperform all baseline schemes.

To the best of our knowledge, this work provides the first rigorous scheduling framework for CC-based streaming over multi-AP WiFi-type networks that explicitly targets fair user goodput, thereby bridging the gap between the theoretical promise of coded caching and its practical deployment in real-world WLAN scenarios such as crowded venues, transportation hubs, and in-flight entertainment systems.

\textbf{Notation:} Vectors are represented by bold letters, and sets are represented by calligraphic letters. $\Bv[i]$ is the $i$th element of vector $\Bv$, and $\CA \backslash \CB$ denotes the set of elements of $\CA$ not in $\CB$. For integer $J$, $[J]$ represents the set $\{1,2,\cdots,J\}$. $\binom{n}{k}$ denotes binomial coefficient, and its value is zero if $n < k$.
We use the symbol $\mathcal{S} \choose k$, where $\mathcal{S}$ is a set, 
to denote the collection of all $|\mathcal{S}| \choose k$ distinct subsets (combinations) 
of size $k$ of the set $\mathcal{S}$.

\section{System Model}
\label{section:sys_model}

\begin{figure}[t]
    %\begin{subfigure}{0.3\columnwidth}
        \centering
        \includegraphics[width = 0.9\columnwidth]{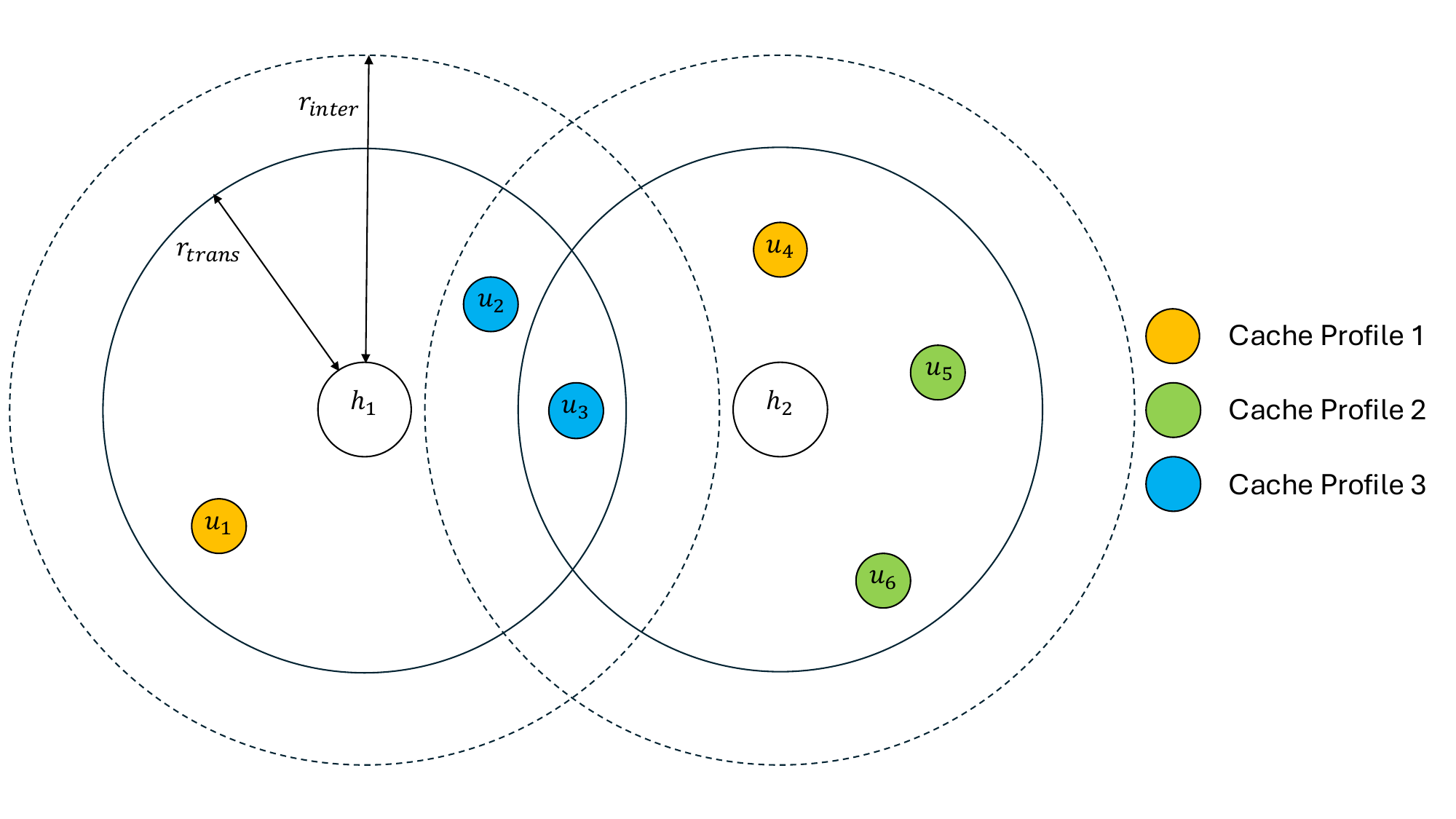} %\label{fig:network_model}
        \vspace{-5pt}
    \caption{Example network with $H=2$, $K=6$, and $L = 3$.}
    %\vspace{-12pt}
    \label{fig:network_model}
\end{figure}

The network model considered in this work comprises a streaming server sending 
on-demand video streams to $K$ wireless users via $H$ APs. 
The server is connected to the APs through an error-free fronthaul network. 
We use $h_i$ and $u_k$, $i \in [H]$ and $k \in [K]$, to denote APs and users, respectively. Fig.~\ref{fig:network_model} illustrates a simple network with $H = 2$ and $K = 6$.
APs have an effective transmission radius of $r_{\mathrm{trans}}$ and an interference radius of $r_{\mathrm{inter}} \ge r_{\mathrm{trans}}$. 
We call an AP $h_i$ {\em active} at a given time slot if it is transmitting data in that slot. 
According to the collision model, denoting the locations of the AP $h_i$ as $\Sfx(h_i)$ and user $u_k$ as $\Sfx(u_k)$ where
$\Sfx(h_i), \Sfx(u_k) \in \mathbb{R}^2$, %respectively, 
$u_k$ can successfully receive the message (packet) sent by $h_i$ if and only if the following two conditions are satisfied:
\begin{enumerate}
    \item $|\Sfx(h_i) - \Sfx(u_k)| \le r_{\mathrm{trans}}$; and
    \item there is no other active AP node $h_{i'}$ such that $|\Sfx(h_{i'}) - \Sfx(u_k)| \le r_{\mathrm{inter}}$.
\end{enumerate}

In video streaming, each video file is split into a sequence of ``chunks'' corresponding to a few seconds of the video. 
%\color{blue}
For example, consider %the example of 
movies of duration $\sim$~90min, and video chunks of duration 
$\sim$~10sec, resulting in a sequence of 540 chunks, whereas at a typical video-coding rate of 2Mb/s, each chunk is 20Mb in size. The solutions proposed in this paper apply at the chunk level.  
%We consider a video library formed by $N$ files, and a per-user cache memory corresponding to $\gamma N $ files, for some $\gamma \in [0,1]$.
We assume that the video library consists of $N$ files, each split into $E$ chunks. Each user can cache $\gamma NE$ chunks in its memory, for some $\gamma \in [0,1]$. %\color{black}
%In general, 
A streaming user generates a sequence of requests for the corresponding chunks 
of its desired video file. %Since 
Users pre-store a portion $\gamma$ of each video chunk, so after each request, only the portion $(1-\gamma)$ of the chunk that is not already in the user's cache must be delivered. 

On-demand streaming sessions are generally asynchronous. This is because synchronous requests from a subset of users require not only that the users stream the same movie, but also that they start streaming within the same time frame of a few seconds (corresponding to the length of each video chunk). 
%since 
%because the number of chunks per video file is very large, the probability that two users request the 
%same chunk in the same time slot is essentially zero, even if the users are streaming the same file. 
This implies that conventional broadcasting, as in live TV where all users ``see the same thing at the same time,'' is not possible in this setting. 

%%%%%%%%%%%%%%%%%%%%%%%%%%%%%%%%%
\section{Coded Caching in Multi-AP Networks}
\label{CC-section}

Following the standard CC setting \cite{maddah2014fundamental}, system operation consists of two phases: placement and delivery. The placement phase is done offline, while delivery takes place 
during the streaming session.\footnote{For example, think of a ``bring-your-own device'' service where users cache segments from a library at home and then stream a particular movie from the library when they join the network (e.g., in a train station, or airport terminal).} 

%%%%%%%%%%%%
\subsection{Placement Phase}

During the placement phase, users' cache memories are filled up with subpackets (i.e., smaller portions) of 
each video chunk, such that a fraction $\gamma$ of each video chunk of each file is cached. 
In order to allow users to join and leave the network and to start their streaming sessions at any time, it is of fundamental importance that the cache placement be completely decentralized. For this purpose, we follow the scheme of \cite{caire,mozhgan} that consists of generating $L$ distinct {\em cache profiles} 
and let each user randomly pick one profile during its offline caching phase. 
The number of cache profiles $L$ is an important system design parameter.
Since the subpacketization is repeated identically for each chunk of each video file, we shall simply denote a generic chunk by $W$ of size $|W|$ bits. 
Assuming that $t = \gamma L$ is an integer,\footnote{If $\gamma L$ is not an integer, the scheme can be modified by cache sharing between two schemes with $t_1 = \lfloor \gamma L \rfloor$ and $t_2 = \lceil \gamma L \rceil$. This is well-known in the CC literature and will not be further discussed here.} 
every chunk $W$ is partitioned into $\binom{L}{t}$ equal-sized subpackets $W_{\CS}$, 
indexed by all possible subsets $\CS \in { [L] \choose t}$. For all $l \in [L]$, the $l$-th cache profile 
$\Phi_l$ is the collection of all subpackets $W_{\CS}$ such that $\CS \ni l$, for all 
chunks of all files in the library. Since each subpacket has size $|W|/{L \choose t}$ bits
and exactly ${L - 1 \choose t - 1}$ subpackets per chunk are included in $\Phi_l$, 
it is immediate to check that the number of cached bits per chunk of each file in the library 
is given by
\[ \frac{|W| {L - 1 \choose t - 1}}{{ L \choose t}} = \frac{|W|t}{L} = |W| \gamma, \]
such that the size of each cache profile $\Phi_l$ is equal to a fraction $\gamma$ 
of the whole library, as required.

We denote by $\SfL(u_k) \in [L]$ the cache profile assignment of $u_k$, i.e., if $\SfL(u_k) = l$ then 
user $u_k$ stores the cache profile $\Phi_l$. 

\begin{exmp}
\label{exmp:placement}
Consider the network in Fig.~\ref{fig:network_model}, which consists of $H=2$ APs and $K=6$ users, with 
$L=3$ cache profiles. Each user is assigned to a cache profile as 
indicated in the figure. 
Suppose that each user can cache $\gamma = 1/3$ of the library, i.e., $t = \gamma L = 1$. 
The cache profiles are obtained by dividing each chunk of every 
file into $\binom{3}{1} = 3$ subpackets. Letting $W$ denote a generic chunk and 
$W_{\{1\}}, W_{\{2\}}, W_{\{3\}}$ denote the corresponding three subpackets, 
cache profile $\Phi_1$ contains all subpackets of type $W_{\{1\}}$ (i.e., all first subpackets of each chunk of each file),  cache profile $\Phi_2$ contains all subpackets of type $W_{\{2\}}$, 
and cache profile $\Phi_3$ contains all subpackets of type $W_{\{3\}}$.
Notice that multiple users may be assigned to the same cache profile, since the assignments are made at random and independently for each user.  For example, here we have that users $u_1,u_4$ are assigned to $\Phi_1$,  $u_5$,$u_6$ to $\Phi_2$, and $u_2$,$u_3$ to $\Phi_3$. 
\end{exmp}

\begin{rem}  \label{uncoded-caching-rem}
    It is important to observe that the conventional ``prefix'' caching \cite{sen1999proxy}, where 
    each user caches the initial $\gamma$-fraction of each chunk, corresponds to a special case of 
    the above scheme with $L = 1$, i.e., where all users have the same cache profile. Therefore, 
    our framework for $L = 1$ encompasses immediately the case of a conventional (uncoded) caching system. 
    \hfill $\lozenge$
\end{rem}

%%%%%%%%%%%%%%%%%%%%%%
\subsection{Delivery Phase}
\label{section_delivery}

During the delivery phase, users request the chunks of their desired video file sequentially. As mentioned earlier, users can start streaming at any time, and once they begin streaming, they repeatedly generate requests for consecutive video chunks. In other words, after receiving a chunk, 
each user immediately generates a new request for the subsequent chunk. 

We assume that a centralized scheduler (run at the server) makes sequential scheduling decisions at each given time slot, 
coinciding with a video chunk duration. 
Notice that the scheduler time slots are generally much longer than
the underlying PHY and MAC data packets since the scheduler operates ``above IP''.  For example, for %a video-coding rate of 
a PHY channel rate of 20Mb/s and chunks of 20Mb, the scheduler 
makes a decision (roughly) every 1 sec. At each slot, given the set of user requests, the scheduler creates {\em codewords} and sends them through the fronthaul to the APs, such that 
a subset of users can receive the codewords via a subset of active APs. 
The determination of which users should receive and which APs should transmit, and the corresponding codeword formation at each time slot, defines a \textit{scheduling decision}. 
Consequently, a sequence of scheduling decisions forms a \textit{scheduling policy} \cite{georgiadis2006resource}.  At each given slot, we define the {\em instantaneous delivery rate} of each user
as the amount of video-coded bits made available to the user's playback buffer in the given slot, normalized by the length $|W|$ of one chunk. 
The instantaneous delivery rate at each given slot depends on the scheduling decision for that slot. We also define the user's goodput as the long-time average of the instantaneous delivery rate over 
a long sequence of slots.  The user's goodput depends on the scheduling policy. 
We shall first clarify these concepts using our running example, and then in full generality. 

\begin{exmp}
\label{exmp:simple_codeword}
    Consider the network in Example~\ref{exmp:placement} and denote the chunks requested by a user $u_k$, $k \in [6]$, as $W_{d_k}$. 
    %by users $u_1$, $u_2$, $u_3$, $u_4$, $u_5$, \textcolor{red}{and $u_6$} at a given time slot as $W_{d_1}, W_{d_2}, W_{d_3}, W_{d_4}$, $W_{d_5}$, \textcolor{red}{and $W_{d_6}$,} respectively. 
    Consider a scheduling decision where AP $h_1$ is idle (no transmission in the current time slot) 
    and users $u_3,u_4,u_5$ are served by AP $h_2$. With the cache placement in Example~\ref{exmp:placement}, the list of cached and requested subpackets by users $u_3-u_5$ is as follows:
    \begin{table}[htb]
        \centering
        \begin{tabular}{c|c|c}
             \textbf{User} & \textbf{Cached} & \textbf{Requested}  \\
             \hline
             $u_3$ & $W_{d_k, \{3\}}, \ \ \forall k \in [6]$ & $W_{d_3, \{1\}}$, $W_{d_3, \{2\}}$ \\
             \hline
             $u_4$ & $W_{d_k, \{1\}}, \ \ \forall k \in [6]$ & $W_{d_4, \{2\}}$, $W_{d_4, \{3\}}$ \\
             \hline
             $u_5$ & $W_{d_k, \{2\}}, \ \ \forall k \in [6]$ & $W_{d_5, \{1\}}$, $W_{d_5, \{3\}}$
        \end{tabular}
        %\caption{Caption}
        %\label{tab:placeholder}
    \end{table}
    \noindent Hence, AP $h_2$ can transmit (in multicast) the codewords 
\begin{equation*}
    \begin{aligned}
X(u_3,u_4) &= W_{d_3,\{1\}} \oplus W_{d_4,\{3\}}, \\ 
X(u_3,u_5) &= W_{d_3,\{2\}} \oplus W_{d_5,\{3\}}, \\  
    X(u_4,u_5) &= W_{d_4,\{2\}} \oplus W_{d_5,\{1\}}, \\
    \end{aligned}
\end{equation*}  
where $\oplus$ denotes the bit-wise XOR operation over the whole subpacket. It is readily seen that 
users $u_3$, $u_4$, and $u_5$ can retrieve their desired subpackets. For example, 
user $u_3$ has $W_{d_4,\{3\}}$ and $W_{d_5,\{3\}}$ in its cache and therefore can 
retrieve $W_{d_3,\{1\}}$ and $W_{d_3,\{2\}}$ by simple XOR-ing the first two codewords, 
and so can also do users $u_4$ and $u_5$. The length of each codeword is equal to the 
length of a subpacket, such that sending the three codewords requires one chunk unit time. 
At the end of this transmission, users $u_3$, $u_4$, and $u_5$ have retrieved 
all the missing subpackets of their requested chunk, i.e., $|W|$ usable bits in their playback buffer,
while users $u_1$ and $u_2$ have not received anything. Therefore, the instantaneous rate vector corresponding to this scheduling decision is $(0,0,1,1,1)$.
\end{exmp}
\subsection{Codeword Formation}
In general, a network with $H$ APs has $2^H - 1$ AP \emph{activation patterns} where at least one AP is active. We denote activation patterns by binary vectors $\Bp_j \in \{0,1\}^H$, $j \in [2^H-1]$.
The activation pattern determines which users can be served by each active AP. This information, combined with users' cache profiles, yields the CC codeword formation. 
In CC, a codeword can be created for every {\em multicast} subset of users $\CT$ of size $0 < |\CT| \le t + 1$, where $\CT$ must be formed by users with {\em distinct} cache profiles.  
Due to multiple choices for $\CT$ for each active AP in a given activation pattern, there may exist multiple candidate codewords for each active AP.

To describe the codeword formation in full generality, we need to introduce some notation. 
For a given activation pattern $\Bp_j$, consider an AP $h_i$ active in $\Bp_j$, i.e., such that $\Bp_j[i] = 1$. 
Denote the set of users that can be served by $h_i$ as $\CU_i$, i.e., $\CU_i$ includes users within distance $r_{\mathrm{trans}}$ of $h_i$ but out of distance $r_{\mathrm{inter}}$ of every other active AP in $\Bp_j$.  
Also, for each $l \in L$, denote by $\CU_i^l$ the subset of users in $\CU_i$ 
assigned to cache profile $\Phi_l$, i.e., the users $u_k \in \CU_i^l$ with $\SfL(u_k) = l$. 
Some of these sets may be empty, and we let $\Sfl(i)$ denote the number of such empty sets. 

To create codewords, we first build a non-empty {\em feasible} 
subset $\CV_i$ of $\CU_i$, where all users in $\CV_i$ have distinct cache profiles. 
Clearly, $|\CV_i| \le L - \Sfl(i)$, and the number of possible ways to build $\CV_i$ is 

\begin{equation}
\bigg(\prod_{l \in [L]} {\left(|\CU_i^l|+1\right)}\bigg) - 1,  \label{possible-ways}
\end{equation}
where addition with one (inside parentheses) is to account for the case no user is selected from $\CU_i^l$, and the subtraction of one is to exclude the empty set.

The users in $\CV_i$ can be served by a number of codewords transmitted by AP $h_i$. 
These codewords are determined as follows: We build the \emph{support} set of $\CV_i$ as 
\begin{equation}
    \CL(\CV_i) = \left\{\SfL(u_k), \forall u_k \in \CV_i \right\},
\end{equation}
and then create its \emph{extended} set $\hat{\CV}_i$ by adding $L - |\CV_i|$ \emph{phantom} users $u_{l}^*$,\footnote{Phantom users are virtual users added only to maintain symmetry in the codeword formation. The same concept is used in~\cite{salehi2020lowcomplexity}.} 
where $\SfL(u_l^*) = l$, for every $l \in [L] \backslash \CL(\CV_i)$. 
Next, a \emph{preliminary} codeword $\hat{X}(\hat{\CT}_i)$ can be built for every multicast 
subset $\hat{\CT}_i$ of $\hat{\CV}_i$ with maximal size $|\hat{\CT}_i| = t+1$ 
as the XOR of subpackets
\begin{equation}
    \hat{X}(\hat{\CT}_i) = \bigoplus_{u_{k} \in {\hat{\CT}_i}} W_{d_{k}, \CL(\hat{\CT}_i) \backslash \{\SfL(u_k)\} } ,
\end{equation}
where 
\begin{equation}
    \CL({\hat{\CT}_i}) = \left\{\SfL(u_k), \forall u_k \in \hat{\CT}_i \right\},
\end{equation}
and $d_k$ is the index of the video chunk requested by $u_k$. 
Finally, the codeword $X(\CT_i)$ is built from $\hat{X}(\hat{\CT}_i)$ by removing the presence 
of phantom users ($\CT_i$ is the multicast set resulting by removing phantom users from $\hat{\CT}_i$).

\begin{exmp}
\label{exmp:codeword_creation}
    Consider again the network in Example~\ref{exmp:placement} in Fig.~\ref{fig:network_model}. 
    In Example~\ref{exmp:simple_codeword}, we provided transmitted codewords for this network when only $h_2$ is active. Now, let us consider a different activation pattern $\Bp_j = [1,0]$, where only $h_1$ is active. 
    For the active AP $h_1$, we have 
    \begin{equation*}
        \CU_1^1 = \{u_1\}, \quad \CU_1^2 = \varnothing , \quad \CU_1^3 = \{u_2,u_3\} .
    \end{equation*}
    %$\CU_2^1 = \{u_4\}$, $\CU_2^2 = \{u_3,u_5\}$, and $\CU_2^3 = \varnothing$. 
    Accordingly, we have five options to choose the feasible set $\CV_1$, given by $\{u_1\}$, $\{u_2\}$, $\{u_3\}$, $\{u_1,u_2\}$ and $\{u_1,u_3\}$. For the choice $\CV_1 = \{u_1,u_2\}$,  since $\SfL(u_1) = 1$ and $\SfL(u_2) = 3$, we have $\CL(\CV_1) = \{1,3\}$. The extended set is obtained by introducing the phantom user $u_2^*$ %with cache profile $l = 2$
    assigned to $\Phi_2$, i.e., $\hat{\CV}_1 = \{u_1,u_2,u_2^*\}$. 
    As a result, we have three subsets $\hat{\CT}_1$ of $\hat{\CV}_1$ with size $t+1 = 2$, resulting in the preliminary codewords
    \begin{equation*}
        \begin{aligned}
            \hat{X}(\{u_1,u_2\}) &= W_{d_1,\{3\}} \oplus W_{d_2,\{1\}}, \\
            \hat{X}(\{u_1,u_2^*\}) &= W_{d_1,\{2\}} \oplus W_{d_{2^*},\{1\}}, \\
            \hat{X}(\{u_2,u_2^*\}) &= W_{d_2,\{2\}} \oplus W_{d_{2^*},\{3\}}. \\
        \end{aligned}
    \end{equation*}
    After removing the effect of the phantom user $u_2^*$, the final codewords are given as
    \begin{equation*}
        \begin{aligned}
            X(\{u_1,u_2\}) &= W_{d_1,\{3\}} \oplus W_{d_2,\{1\}}, \\
            X(\{u_1\}) &= W_{d_1,\{2\}}, \\
            X(\{u_2\}) &= W_{d_2,\{2\}}.
        \end{aligned}
    \end{equation*} 
\end{exmp}

\subsection{Instantaneous Rate Calculation}
%In general, 
At each scheduling slot, the scheduler chooses an AP activation pattern $\Bp_j$ and %some 
suitable codewords as described above. We let $\Sfc(\Bp_j)$ denote the number of 
possible codewords for a given activation pattern $\Bp_j$. Formally, a scheduling decision $(j,s)$ corresponds to the selection of activation pattern $\Bp_j$ and codeword choice $s \in [\Sfc(\Bp_j)]$.
Let us consider an activation pattern $\Bp_j$ and an active AP $h_i$ in $\Bp_j$. 
It can be easily verified that the codeword formation process described above guarantees that for a given $\CV_i$, 
each user $u_k \in \CV_i$ obtains its requested chunk after 
\begin{equation}
\Sfn(\CV_i) = {\binom{L}{t+1} - \binom{L-|\CV_i|}{t+1}}  \label{eq:numberofslots}
\end{equation}
codeword transmissions, where each transmission takes the length of one subpacket, i.e.,  $|W|/\binom{L}{t}$ 
bits. Hence, the instantaneous delivery rate of each user $u_k \in \CV_i$ is given by  
\begin{equation}
r(\CV_i) = \frac{|W|}{\Sfn(\CV_i) \frac{|W|}{{L \choose t}}} = \frac{{L \choose t}}{\Sfn(\CV_i)}.  \label{inst-rate-expression}
\end{equation}
Note that $r(\CV_i)$ in \eqref{inst-rate-expression} is calculated according to the fact that $\binom{L-|\CV_i|}{t+1}$  preliminary codewords $\hat{X}(\hat{\CT}_i)$ are ignored, i.e., not transmitted,  
as they include only phantom users. %\textcolor{red}{In addition, for fixed $L$ and $t$, $r(\CV_i)$ depends only on the size of the set $\CV_i$, so, from now on, we will use $r(\CV_i)$ and $r(|\CV_i|)$ interchangeably.}

Now, to find the instantaneous rate vector $\Br(j,s)$ containing the instantaneous rates of all users in the network, we need to: 
%\begin{enumerate}
1) for each active AP $\{h_i : \Bp_j[i] = 1\}$, choose the feasible set $\CV_i$; 
2) for each $u_k \in \CV_i$, set the element $k$ of the vector $\Br(j,s)$ equal to $r(\CV_i)$ as in \eqref{inst-rate-expression}; 
3) fill other elements of $\Br(j,s)$ with zeros.
%\end{enumerate}
\begin{rem}
\label{rem:cardinality}
    Notice that for fixed $L$ and $t$, $r(\CV_i)$ depends only on %the size of the set 
    $|\CV_i|$; it decreases as $|\CV_i|$ is increased up to $|\CV_i| = L-t$, but then remains fixed when $L-t \le |\CV_i| \le L$. This is an important feature which will be exploited when we develop the reduced-complexity dynamic solution in Section~\ref{sec:new_dynamic}. 
    %(We haven't started explicitly talking about scheduling and maximal vectors yet, so this part should be sufficient here.)
    \hfill $\lozenge$
\end{rem}
%\todo[inline]{a remark here on how the rate values change can help later with the dynamic algorithm.}
\begin{exmp}
\label{exmp:rate_vector}
    Consider the same network in Example~\ref{exmp:placement}, and assume the activation pattern is $\Bp_j = [1,0]$. For active AP $h_1$ with $\CV_1 = \{u_1,u_2\}$, we get the three codewords shown in Example~\ref{exmp:codeword_creation}. The transmission of these three codewords requires 
    one slot and delivers one chunk to users $u_1$ and $u_2$ and nothing to the other users, yielding the instantaneous rate vector $(1,1,0,0,0)$. 
    Similarly, it can be seen that for $\CV_1 = \{u_1\}$, we get two codewords 
    \begin{equation*}
        X(\{u_1\}) = W_{d_1,\{2\}}, \quad X(\{u_1\}) = W_{d_1,\{3\}}.
    \end{equation*}
    The transmission of these two codewords requires $2/3$ of a slot and delivers one chunk to user $u_1$, yielding the instantaneous rate vector $(\frac{3}{2},0,0,0,0)$.
\end{exmp}

%%%%%%%%%%%%%%%%%%%%%%%%%%%%%%%%%%%%%%%%%%%%%%%%%%%%%%%%%%%%%%%%%%%%%%%%%%%%%
\section{Fairness Scheduling}
\label{scheduling}

As discussed in Section~\ref{CC-section}, each scheduling decision $(j,s)$ results in an instantaneous rate vector $\Br(j,s)$, where the $k$-th element, $k \in [K]$, is the number of useful 
bits 
%delivered 
made available to user $u_k$ in the scheduling slot under decision $(j,s)$, normalized by the chunk size.  
%As anticipated before, 
A scheduling policy consists of a sequence of scheduling decisions $\{(j_t, s_t) : t = 1, 2, \ldots\}$, whereas the goodput $K$-tuple (a vector in $\mathbb{R}_+^K$) achieved by a given scheduling policy is given by the long-term time average
\begin{equation}
\bar{\Br} = \lim_{T \rightarrow \infty} \frac{1}{T} \sum_{t = 1}^T \Br(j_t,s_t),
\end{equation}
if such a limit exists. It is well-known~\cite{georgiadis2006resource}
that the goodput region of the network is given by the convex hull of all instantaneous 
rate vectors, i.e., 
\begin{equation} 
{\cal R} = \mathrm{Conv}\left(\Br(j,s) : j \in [2^H], s \in [\Sfc(\Bp_j)]\right).
\end{equation}
Since the number of instantaneous rate vectors is finite, ${\cal R}$ is a convex polytope in $\mathbb{R}_+^K$. 
Furthermore, %it is also well-known that 
any point in the goodput region can be achieved by some 
randomized stationary scheduling policy, i.e., a policy that at each slot chooses (independently over the time slots) decision $(j, s)$ with probability $\pi(j,s)$~\cite{georgiadis2006resource}.

In general, we are interested in scheduling policies that operate the system on the Pareto boundary 
of ${\cal R}$. However, the choice of the operating point may correspond to different criteria of optimality. Following~\cite{georgiadis2006resource}, the operating point of the scheduler is determined
as the point of the goodput region that maximizes a suitable 
{\em network utility function} of the user goodputs. By choosing a concave componentwise non-decreasing network utility function, one can impose a desired fairness criterion. For example, we may choose any function from the well-known and widely used $\alpha$-fairness family~\cite{fairness}. In particular, for $\alpha = 1$ the corresponding fairness criterion is referred to as {\em proportional fairness}, 
for $\alpha = 0$, the scheduler maximizes the sum over the goodputs (generally unfair), 
and for $\alpha \rightarrow \infty$, we obtain max-min (or ``hard'') fairness. %\textcolor{red}{To be able to stream, user goodputs must be above the video-coding rate, and the relevant utility function that can ensure this is hard fairness, since it maximizes the minimum user goodput. However, if proportional fairness is employed, at the expense of decrease of the goodputs of some users, most users can stream at significantly higher quality when adaptive quality coding is used~\cite{lefteris-adaptivequality}. 
%So, in this paper,} 
%\textcolor{red}{To enable continuous streaming, user goodputs must exceed the video-coding rate~\cite{bethanabhotla2014adaptive,caire_videobuffer}. The utility function that directly enforces this requirement is 
%hard fairness, which maximizes the minimum user goodput. However, when proportional fairness is employed, at the expense of a reduction in the goodputs of some users, most users can stream at significantly higher quality when adaptive quality coding is used~\cite{lefteris-adaptivequality}. So in this paper,} 
We focus on proportional fairness and hard fairness, corresponding to the maximization of the 
network utility functions
\begin{equation}
\label{eq:utility_function}
\Sff_{\mathrm{pf}}(\bar{\Br}) \coloneqq \sum_k \log(\bar{\Br}[k])
\end{equation}
for proportional fairness (PF), and 
\begin{equation}
\label{eq:hard_fairness_function} 
\Sff_{\mathrm{hf}}(\bar{\Br}) \coloneqq \min_{k} \bar{\Br}[k] 
\end{equation}
for hard fairness (HF).
\begin{rem}
\label{remark:geom_mean}
Maximizing the PF function is, in fact, equivalent to maximizing the geometric mean of the 
users' goodput vector. This is because we have
\begin{equation}
\frac{1}{K}\sum_k \log(\bar{\Br}[k])=\log\left(\prod_k(\bar{\Br}[k])\right)^{\frac{1}{K}}.
\end{equation}
In this regard, we shall present PF results in terms of the goodput geometric mean to compare the proposed schemes in Section~\ref{section:sim_results}. %since this has the meaning of a rate in chunks per unit time.
\hfill $\lozenge$
\end{rem}
%%%%%%%%%%%%%%%%%%%
%%%%%%%%%%%%%%%%%%%
\subsection{The Optimization Problem}
\label{section:opt-problem}
%\hspace{-5pt}
The maximization of the network utility function and the determination of the corresponding optimum goodput point in ${\cal R}$, can be concisely stated as 
\begin{equation}
\label{eq:optimization_problem_1}
    \begin{aligned}
   \arg     \max \Sff(\bar{\Br}) , \quad & \Sff\in{\lbrace\Sff_{\mathrm{pf}},\Sff_{\mathrm{hf}}\rbrace}, 
        \\
        s.t. \qquad \bar{\Br} \in {\cal R}. 
    \end{aligned}
\end{equation}
Since the objective functions are concave, \eqref{eq:optimization_problem_1} is a convex problem. 
However, it is generally difficult to describe the convex polytope ${\cal R}$ in terms of 
linear inequalities (i.e., its supporting hyperplanes).  
Since each point of ${\cal R}$ is achievable by a stationary policy, Problem \eqref{eq:optimization_problem_1}
can be restated in an equivalent form in terms of the probabilities 
$\{\pi(j,s)\}$ that define a stationary policy. This yields 
\begin{equation}
\label{eq:optimization_problem_main}
    \begin{aligned}
      \arg  \max \;\;  \Sff(\bar{\Br}) , \quad & \Sff\in{\lbrace\Sff_{\mathrm{pf}},\Sff_{\mathrm{hf}}\rbrace}, %&= \sum_{k \in [K]} \log(\bar{\Br}[k]) 
        \\
        s.t. \qquad \bar{\Br} = \sum_{j \in [2^H-1]} & \sum_{s \in [\Sfc(\Bp_j)]} \pi(j,s) \Br(j,s), \\
         \pi(j,s)\geq{0}, \quad & \sum_{j \in [2^H-1]} \sum_{s \in [\Sfc(\Bp_j)]} \pi(j,s) = 1 . 
    \end{aligned}
\end{equation}
Notice that this problem 
%and the problem in \eqref{eq:optimization_problem_main} have essentially the same complexity 
%in the sense that they 
requires the enumeration of all (maximal) instantaneous rate vectors. %~\footnote{
A maximal vector in a set of vectors in $\mathbb{R}^K$ is a vector that is not componentwise dominated by any other vector in the set. It is clear that %search in \eqref{instantaneous-wsrm} and 
the sum over $(j,s)$ in \eqref{eq:optimization_problem_main} can be restricted to maximal instantaneous rate vectors without loss of optimality. %} 
However, this is generally a very hard task unless the network is very small.  In fact, there exist $2^H-1$ activation patterns, and the number of scheduling decisions per active AP given an activation pattern is given by~\eqref{possible-ways}. This means that, the number of scheduling decisions $(j,s)$, i.e., variables $\pi(j,s)$ in~\eqref{eq:optimization_problem_main} is approximately given by 
\begin{equation}
\label{eq:complexity_calc_opt}
    \sum_{j \in [2^H-1]}\prod_{\substack{i \in [H],\\\Bp_j[i] = 1}}
    \prod_{l \in [L]} {(|\CU_i^l|+1)}.
\end{equation}
For fixed $L$, this number grows faster than exponentially when $H$ and the number of users inside the AP transmission radii increase, i.e., when $|\CU_i^l|$ increases for each $i$ and $l$. As a quick example, for a network of only $H=4$ APs and $L=4$, %if there are even only 
even for just $2$ user from each cache profile and a total of $8$ users that can be served by each AP simultaneously,  %just for 
the single activation pattern $\Bp_j=(1,1,1,1)$ yields a number of scheduling decisions 
equal to $3^{16}\approx O(10^7)$.

\begin{rem}
    \label{rem:complexity_remark}
    In fact, the complexity order in~\eqref{eq:complexity_calc_opt} is an upper bound to the real complexity value. This is because, it can be inferred from Remark~\ref{rem:cardinality} %that the feasible sets $\CV_i$ with $\CV_i$ 
    that for a given activation pattern $\Bp_j$ and active AP $h_i$, some feasible sets do not need to be considered when forming the codeword, i.e., 
    %that 
    if $L-\Sfl(i)>L-t$, one does not need to consider the feasible sets $\CV_i$ with $L-t \le |\CV_i| < L-\Sfl(i)$, as the users' individual rates won't change in this region and the corresponding rate vector will be dominated by the rate vector using all the $L-\Sfl(i)$ users in codeword formation. 
    %((I wrote again here lik this, since I think otherwise it would be confusing to write/understand why this happens without explicitly stating it mathematically.))
    %some feasible sets do not need to be considered when forming the codeword, because their corresponding rate vector would be dominated by another rate vector for  $\CV_i$ with $L-t \le |\CV_i| \le L$ only the
    %be seen from Remark~\ref{rem:cardinality} that not every increase in the size of the feasible set would give a lower rate, so the rate vectors corresponding to feasible sets with the biggest size would dominate the other
    Nevertheless, even after applying this reduction, the general scaling law of the complexity order will remain unchanged,
    %; the right-hand side of the first summation will still be exponential, 
    and solving the optimization problem will still be infeasible for large networks. %Furthermore, some activation patterns where only a small set of APs are active may not need to be considered if activating more APs doesn't create 
    Furthermore, certain activation patterns in which only a small subset of APs is active need not be considered if activating additional APs does not introduce any interference among all the users that can be chosen for the codeword formation. For instance, if the interference radii of $h_1$ and $h_2$ %don't intersect and are far apart, it is redundant to obtain the instantaneous rate vectors of the cases where only 
    do not overlap and are sufficiently separated, it is redundant to compute instantaneous rate vectors corresponding to cases in which only one of the two APs is active, %because any such rate vector will be dominated by another rate vector obtained by activating both APs. However, the number of activation patterns that can be discarded, if any, depends on the specific network topology.
    since any such rate vector is dominated by a rate vector obtained when both APs are activated. However, the set of activation patterns that can be discarded, if any, depends on the specific network topology.
    \hfill $\lozenge$
\end{rem}

\subsection{Complexity Reduction}
%\label{complexity-reduction}

The complexity of the optimization problem~\eqref{eq:optimization_problem_main} can be reduced by noticing that some users in the network are \textit{equivalent}. Two users are \textit{equivalent} if they have the same cache profile and the same spatial properties, i.e., if both of them are within the transmission and interference radii of the same APs. It is easy to see that if two users are 
equivalent, then the instantaneous rate vectors are symmetric in such users in the following sense: 
suppose that $u_k$ and $u_{\tilde{k}}$ are equivalent users and consider an instantaneous rate vector
$\Br$ with $k$-th element $\Br[k] > 0$ (clearly, $\Br[\Tilde{k}]=0$ as $u_k$ and $u_{\tilde{k}}$ have the same cache profile and therefore cannot be served together in the same multicast group). Then, there exists another instantaneous rate vector 
$\Tilde{\Br}$ such that $\Tilde{\Br}[\Tilde{k}] = \Br[k]$, $\Tilde{\Br}[k]=0$, and 
$\Tilde{\Br}[k']=\Br[k']$ for all $k'\neq k,\Tilde{k}$. 
%Furthermore, it must be 
%$\Br[\Tilde{k}]=\Tilde{\Br}[k]=0$ since $u_k$ and $u_{\tilde{k}}$ have the same cache profile and therefore cannot be served together in the same multicast group. 
Furthermore, it can be easily verified that the solution of the optimal fairness point problem~\eqref{eq:optimization_problem_main} assigns equal probability to the rate vectors $\Br$ and $\tilde{\Br}$.

%Some complexity reduction can be obtained 
By grouping users into equivalent classes with respect to the above defined equivalence relation (notice that some equivalent classes are indeed singletons, 
since there may be users that are not equivalent to any other user), %and considering the 
%reduced 
the complexity of the network can be reduced %obtained 
by retaining only one representative user for each equivalent class. 
After computing the set of instantaneous rate vectors for the reduced network, the set of rate vectors 
for the original network can be obtained simply by expanding each class 
representative rate entries into a uniform (equal entries) subvector, of length equal to the size of the
corresponding equivalent class, and entries that sum up to the rate of the representative. 
For example, if user $u_k$ is the representative of a class of (say) $\ell$ users and, for a given 
scheduling decision $(j,s)$ in the reduced network, receives an instantaneous rate of $r_k$, 
the $\ell$ elements of the rate vector of the original network corresponding to the equivalent class will all be equal to $r_k/\ell$. 

The equivalent user approach yields some moderate complexity reduction with respect to the direct formulation 
of problem~\eqref{eq:optimization_problem_main}. Nevertheless, it remains unsuitable for scaling to large networks.

\begin{exmp}
\label{ex:equivalent users}
Consider the network in Fig.~\ref{fig:network_model}. %, but let $u_3$ have cache profile $1$ instead of $2$, and $u_4$ have cache profile $2$ instead of $1$.
For this network, we have put the maximal instantaneous rate vectors and the results of the optimization problem in~\eqref{eq:optimization_problem_main} for both PF and HF in Table~\ref{tab:equivusertable}. The first column in Table~\ref{tab:equivusertable} includes 
the case where equivalent users are not considered. %data for the case where all rate vectors (without considering equivalent users) are considered. 
As can be seen, there exists a total number of 11 maximal instantaneous rate vectors in this case.

However, %it can be easily checked that $u_5$ and $u_6$ are equivalent: from Fig (fig), they are both assigned with the same cache profile and are served only by $h_2$, and from Table (tb), they have symmetric rate values, i.e., [COMPLETE]. 
it can be easily verified from Fig~\ref{fig:network_model} that $u_5$ and $u_6$ are equivalent users, as they %both have cache  profile $2$ 
are both assigned to cache profile $\Phi_2$, are within the transmission radius of $h_2$, and outside of the transmission radius of $h_1$. %As a result, 
This is also evident from the first column of Table~\ref{tab:equivusertable}, as $u_5$ and $u_6$ are symmetric in instantaneous rate vectors $\{\Br(2,1), \Br(2,2)\}$, $\{\Br(3,1), \Br(3,2)\}$, and $\{\Br(3,4), \Br(3,5)\}$. As a result, one can %represent both of them with an equivalent user, 
retain one of them as the class representative user, and calculate the rate vectors for the
reduced network, as shown in the second column of Table~\ref{tab:equivusertable}. By expanding the rate vectors and solving the optimization problem, as shown in the third column of Table~\ref{tab:equivusertable}, final rate and fairness function values remain unchanged, while the number of maximal rate vectors is reduced from $11$ to $8$. As an extra note, we notice also from Table~\ref{tab:equivusertable} that since $u_5$ and $u_6$ share the network resources equally,  $\pi_{pf}(2,1)=\pi_{pf}(2,2)=\frac{1}{2}\tilde{\pi}_{pf}(2,1)$ and $\pi_{pf}(3,1)=\pi_{pf}(3,2)=\frac{1}{2}\tilde{\pi}_{pf}(3,1)$, and a similar pattern also holds for HF. 
\begin{table*}[t]
%\label{tab:equivusertable}
    \centering
    \begin{tabular}{>{\centering\arraybackslash}m{0.15\textwidth}!{\vrule width 1pt}>{\centering\arraybackslash}p{0.25\textwidth}|>{\centering\arraybackslash}p{0.25\textwidth}|>{\centering\arraybackslash}p{0.25\textwidth}}
         \textbf{Activation pattern} & \textbf{Original rate vectors} & \makecell{\textbf{Reduced rate vectors}  \\ ($u_5$ and $u_6$ merge)} & \makecell{\textbf{New rate vectors} \\ (generated from reduced ones)}  \\
         \Xhline{1pt}
         $\Bp_1=(1,0)$ & 
    $\begin{aligned}
        \Br(1,1)&=(1,1,0,0,0,0), \\ 
        \Br(1,2)&=(1,0,1,0,0,0), \\
        \Br(1,3)&=(0,1.5,0,0,0,0)& \\ 
        \Br(1,4)&=(0,0,1.5,0,0,0)
    \end{aligned}$ &
    $\begin{aligned}
        \tilde{\Br}_{\mathrm{red}}(1,1) &= (1,1,0,0,0) \\
        \tilde{\Br}_{\mathrm{red}}(1,2) &= (1,0,1,0,0) \\
        \tilde{\Br}_{\mathrm{red}}(1,3) &= (0,1.5,0,0,0) \\
        \tilde{\Br}_{\mathrm{red}}(1,4) &= (0,0,1.5,0,0)
    \end{aligned}$ &
    $\begin{aligned}
        \tilde{\Br}(1,1)&=(1,1,0,0,0,0) \\
        \tilde{\Br}(1,2)&=(1,0,1,0,0,0) \\
        \tilde{\Br}(1,3)&=(0,1.5,0,0,0,0) \\
        \tilde{\Br}(1,4)&=(0,0,1.5,0,0,0)
    \end{aligned}$ \\
         \hline
         $\Bp_2=(0,1)$ &
    $\begin{aligned}
        \Br(2,1)&=(0,0,1,1,1,0) \\
        \Br(2,2)&=(0,0,1,1,0,1)
    \end{aligned}$ &
    $\begin{aligned}
        \tilde{\Br}_{\mathrm{red}}(2,1) &= (0,0,1,1,1)
    \end{aligned}$ &
    $\begin{aligned}
        \tilde{\Br}(2,1)&=(0,0,1,1,0.5,0.5)
    \end{aligned}$ \\
         \hline
         $\Bp_3=(1,1)$ & 
    $\begin{aligned}
        \Br(3,1)&=(1.5,0,0,1,1,0)  \\ 
        \Br(3,2)&=(1.5,0,0,1,0,1) \\
        \Br(3,3)&=(1.5,0,0,1.5,0,0) \\
        \Br(3,4)&=(1.5,0,0,0,1.5,0) \\
        \Br(3,5)&=(1.5,0,0,0,0,1.5)
    \end{aligned}$ &
    $\begin{aligned}
        \tilde{\Br}_{\mathrm{red}}(3,1) &= (1.5,0,0,1,1) \\
        \tilde{\Br}_{\mathrm{red}}(3,2) &= (1.5,0,0,1.5,0) \\
        \tilde{\Br}_{\mathrm{red}}(3,3) &= (1.5,0,0,0,1.5)
    \end{aligned}$ &
    $\begin{aligned}
        \tilde{\Br}(3,1)&=(1.5,0,0,1,0.5,0.5) \\
        \tilde{\Br}(3,2)&=(1.5,0,0,1.5,0,0) \\
        \tilde{\Br}(3,3)&=(1.5,0,0,0,75,0.75)
    \end{aligned}$ \\
    \Xhline{1pt}
    \textbf{Non-zero probabilities from~\eqref{eq:optimization_problem_main} for PF} &
    $\begin{aligned}
    &\pi_{pf}(1,3)\approx0.1667 \\
    &\pi_{pf}(2,1)=\pi_{pf}(2,2)\approx0.2083 \\
    &\pi_{pf}(3,1)=\pi_{pf}(3,2)\approx0.2083
    \end{aligned}$ & 
    $-$ &
    $\begin{aligned}
        &\tilde{\pi}_{pf}(1,3)\approx0.1667 \\
        &\tilde{\pi}_{pf}(2,1)\approx0.4167 \\
        &\tilde{\pi}_{pf}(3,1)\approx0.4166
    \end{aligned}$\\
    \hline
    \textbf{PF goodput vector} &
    $\begin{aligned}
        &\bar{\Br}_{pf}^\star \approx \\ (0.62,0.25,&0.42,0.83,0.42,0.42)
    \end{aligned}$ 
    %(0.62,0.25,0.42,0.83,0.42,0.42) 
    & 
    $-$ &
    $\begin{aligned}
        &\bar{\Br}_{pf}^\star \approx \\ (0.62,0.25,&0.42,0.83,0.42,0.42)
    \end{aligned}$
    \\
    \hline
    \textbf{PF geometric mean} & $0.46$ & $-$ & $0.46$ \\
    \Xhline{1pt}
    \textbf{Non-zero probabilities from~\eqref{eq:optimization_problem_main} for HF} & $\begin{aligned}
    &\pi_{hf}(1,3)\approx0.2857 \\
    &\pi_{hf}(2,1)=\pi(2,2)\approx0.2143 \\
    &\pi_{hf}(3,4)=\pi(3,5)\approx0.1429
    \end{aligned}$
    &
    $-$ &
    $\begin{aligned}
        \tilde{\pi}_{hf}(1,3)\approx0.2857\\
        \tilde{\pi}_{hf}(2,1)\approx0.4286 \\
        \tilde{\pi}_{hf}(3,3)\approx0.2857
    \end{aligned}$ \\
    \hline
    \textbf{HF goodput vector} & $\begin{aligned}
        &\bar{\Br}_{hf}^\star \approx\\ (0.43,0.43,&0.43,0.43,0.43,0.43)
    \end{aligned}$
    &
    $-$ &
    $\begin{aligned}
        &\bar{\Br}_{hf}^\star \approx\\ (0.43,0.43,&0.43,0.43,0.43,0.43)
    \end{aligned}$ \\
    \hline
    \textbf{HF minimum rate} & $ %\Sff_{\mathrm{hf}}(\bar{\Br}_{hf}^\star) = 
    0.43$ & $-$ & $%\Sff_{\mathrm{hf}}(\bar{\Br}_{hf}^\star) = 
    0.43$ \\
    \end{tabular}
   % \hspace{3pt}
    \caption{The maximal instantaneous rate vectors and the solution of~\eqref{eq:optimization_problem_main} for the network in Fig~\ref{fig:network_model}, w/ and w/o considering the equivalent users.}
    \label{tab:equivusertable}
    %\hspace{-5pt}
\end{table*}
\end{exmp}
%\todo[inline]{A new section could be put here.}
%A systematic way to generate lower complexity heuristic (suboptimal) solutions consists of
%restricting the number of instantaneous rate vectors, thus reducing in a controlled way the goodput region 
%${\cal R}$. %Since each instantaneous rate vector corresponds to a scheduling decision, in terms of 
%AP activation pattern and CC codewords, this in turns yields a systematic way to 
%restrict the possible scheduling decisions. 
%where $\Sff_{\mathrm{pf}}=\sum_{k \in [K]} \log(\bar{\Br}[k])$ and $\Sff_{\mathrm{hf}}=\min_{k \in [K]} %\bar{\Br}[k]$.
%%%%%%%%%%%%%%%%%%%
%%%%%%%%%%%%%%%%%%%
\section{Dynamic Scheduling Solutions}
\label{section:dynamic}

The solution to~\eqref{eq:optimization_problem_main} yields a stationary scheduling policy that
maximizes the desired network utility function. However, the resulting scheduler is ``static'', in the sense that it will choose decision $(j,s)$ with probability $\pi(j,s)$ irrespective of the actual status of chunk delivery to the users. Hence, optimality is approached only in the limit of a very long averaging time, and the actual sample paths of the delivery process for any given user (i.e., the number of delivered chunks per unit time averaged over a short window of slots) may actually deviate significantly from the long-term average. 
In addition, such a policy must be recalculated as the network topology changes, for example, when new users start streaming and/or when currently streaming users stop. 
Using the well-known theory developed in \cite{georgiadis2006resource,queue2,queue_exact}, 
a dynamic policy based on the Lyapunov DPP method 
can be readily obtained by defining a vector of 
recursively updated priority weights $\BQ_t \in \mathbb{R}_+^K$, usually referred to as 
{\em virtual queue backlogs}, with update equation
\begin{equation}
\label{queue_eq}
    \BQ_{t+1}=[\BQ_t - \BR_t]_+ + \BA_t,  
\end{equation}
where $[\cdot]_+$ indicates the (componentwise) positive part, 
$\BR_t$ is an instantaneous rate vector, and $\BA_t$ is the vector of {\em 
virtual arrival processes} at slot time $t$.  The DPP scheduler iterates the following three steps:
\begin{enumerate}
\item Virtual arrival determination as the solution of the auxiliary optimization problem:
\begin{equation}
\BA_t = \arg\max_{\mathbf{a} \in [0, A_{\max}]^K} \;\; V \Sff(\mathbf{a}) - \mathbf{a}^{\sf T}  \BQ_t, \label{virtual_arrival}
\end{equation}
where $V > 0$ and $A_{\max} > 0$ are two control parameters.
\item Instantaneous rate determination as the solution of the weighted sum-rate maximization (wsrm) problem
\begin{equation} 
\BR_t = \arg\max_{\Br \in \{\Br(j,s)\}} \; \Br^{\sf T} \BQ_t. \label{instantaneous-wsrm}
\end{equation}
\item Update of the virtual queues according to \eqref{queue_eq}.
\end{enumerate}
The scheduler is initialized with a suitable initial backlog value, e.g., $\BQ_0 = {\bf 0}$. It can be shown that 
for a stationary network (in our case, for fixed topology and user streaming activity), 
the long term average $\bar{\Br}^{\rm dpp}$ of the instantaneous rates produced by the DPP algorithm satisfies 
the optimality condition
\begin{equation}
\label{eq:dpp-convergence}
    \Sff(\bar{\Br}^{\rm dpp}) \geq \Sff(\bar{\Br}^\star) - O(1/V)
\end{equation}
where $\bar{\Br}^\star$ is the solution of \eqref{eq:optimization_problem_1} and $O(1/V)$ is a term
that decreases as $1/V$, provided that the hypercube $[0, A_{\max}]^K$ contains ${\cal R}$. In our case, this condition is trivially satisfied by noticing from \eqref{inst-rate-expression} that the instantaneous rate of any user cannot be larger than ${L \choose t}$, therefore ${\cal R}$ is necessarily contained in the hypercube
$[0, A_{\max}]^K$ for $A_{\max}={L \choose t}$.

%%%%%%%%%%%%%%%%%%%%%%%%%%%%%%%%%%%%%
%The advantage of the DPP scheduler is that it can adapts to changing topologies since the virtual 
%queues keep track of the state of the past chunk delivery process. Of course, in the case of 
%time-varying network topologies, the optimality result above does not hold since the optimal solution
%is undefined. In fact, for each different topology we have a different region ${\cal R}$ and 
%therefore a different solution of \eqref{eq:optimization_problem_1}. In this case, the value of 
%$V$ can be determined by seeking a tradeoff between ability to adapt quickly, and ``steady state'' 
%near-optimality performance. 
%%%%%%%%%%%%%%%%%%%%%%%%%%%%%%%%%%%%%

The auxiliary problem \eqref{virtual_arrival} is convex and easily solved in closed form for the class of
$\alpha$-fairness functions $\Sff$. In particular, for 
$\Sff(\cdot) = \Sff_{\mathrm{pf}}(\cdot)$ the solution of \eqref{virtual_arrival} is given by 
\begin{equation}
    \label{arrival_simple_eq}
    \BA_{t}[k]=\min\left(\frac{V}{\BQ_{t}[k]},A_{\max}\right), 
\end{equation}
and for $\Sff(\cdot) = \Sff_{\mathrm{hf}}(\cdot)$, the solution is given by 
\begin{equation}
    \label{eq:hf_arrival}
    \BA_{t}[k]=
    \begin{cases}
    A_{\max} & \text{if $V > \sum_k\BQ_t[k]$}, \\
    0 & \text{otherwise}.
    \end{cases}
\end{equation}
In contrast, the instantaneous %weighted sum rate maximization 
wsrm in
\eqref{instantaneous-wsrm} is a very hard combinatorial problem since we need to search over all possible
(discrete) maximal instantaneous rate vectors $\{\Br(j,s)\}$. 

%While this general approach to systematically generate heuristic solutions and therefore scheduling policies
%yields potentially a large variety of possibilities, which may be targeted for particular network topologies, 
%here we explore one method that yields a near-optimal operating point when $L$ becomes large, but it is yet too complex for large networks, and one method based on virtual queues, which is instead full scalable but somehow suboptimal. 

\subsection{Reduced-Complexity Dynamic Solution}
\label{sec:new_dynamic}

One main source of complexity in solving the wsrm \eqref{instantaneous-wsrm}
is that, for each AP activation pattern, a typically large number of scheduling decisions (i.e., options for CC codeword formation) exist. 
Here, we consider an approach that constructs a reduced set of candidate instantaneous rate vectors containing the optimal solution of 
the wsrm by taking into account the properties of users with the same cache profile served by the same active AP and their queue weights in $\BQ_t$. 
We refer to this reduced complexity solution of \eqref{instantaneous-wsrm} as ``dynamic'' since the construction of the reduced set of candidate instantaneous rate vectors is based on the queue weights $\BQ_t$, which is dynamically updated via \eqref{queue_eq}.

Consider an activation pattern $\Bp_j$ and a subset of users $\CK \subseteq [K]$, where all users in $\CK$ are assigned the same cache profile and can be served by the same active AP $h_i$ in $\Bp_j$. %Consider such a subset $\CK$ for which all users can be served by the active AP $h_i$. 
For this subset, the following properties hold:
\begin{itemize}
    \item 
    the users in $\CK$ cannot receive simultaneous transmissions from $h_i$, so only one user in the subset can attain a nonzero rate in any feasible instantaneous rate vector, and
    \item 
    since user rates are multiplied by their virtual queue backlogs in the objective function of~\eqref{instantaneous-wsrm}, it is sufficient to consider only the user with the largest queue backlog in $\CK$ when forming the codeword.
\end{itemize}

%Before continuing with the algorithm description, we need a new definition. A subset of users is called equivalent with respect to a given activation pattern $\Bp_j$ if all of them can be served by the same AP and are assigned to the same cache profile. Let us consider such a subset of users, where all of them can be served by the active AP $h_i$. For this subset of users, the following properties hold: 1) these users can't receive simultaneous transmission from $h_i$, so only one of them can have a nonzero rate in an instantaneous rate vector, 2) since user rates are multiplied with their virtual queue backlogs in the weighted sum-rate maximization in~\eqref{instantaneous-wsrm}, it is sufficient to consider only the user with the largest queue backlog in the codeword formation.

Now, to reduce the number of instantaneous rate vectors for a given activation pattern $\Bp_j$, we proceed as follows:

1) For each active AP $h_i$ and each cache profile $l$ where $\CU_i^l$ is nonempty, find the user with the largest queue backlog. Let $\CC_i$ denote the set of such users for all cache profiles. Clearly, $|\CC_i| = L-\Sfl(i)$, since there are $\Sfl(i)$ cache profiles where $\CU_i^l$ is empty. 
    
2) %For the users selected in Step 1,
Consider only the users in $\{\CC_i\}_i$ while forming the codewords, and find the instantaneous rate vector maximizing the weighted sum-rate for the activation pattern $\Bp_j$ by identifying and combining the multicast groups that yield the highest weighted sum-rate for each active AP.
%By considering only the users selected in Step 1 when forming the codeword, we find the instantaneous rate vector that maximizes the weighted sum rate for the activation pattern $\Bp_j$. First, we identify the multicast groups that yield the highest weighted-sum rate for each active AP, and then combine them to obtain the instantaneous rate vector. 

Notice that for a given active AP $h_i$, %since 
each of the $L-\Sfl(i)$ users in $\CC_i$ has a different cache profile, so one can create a multicast group, i.e., a feasible set, by using any combination of these users. However, from Remark~\ref{rem:cardinality} %\color{blue}rem\color{black}
, we know that the instantaneous rate of the users in a feasible set $\CV_i$ depends only on
$|\CV_i|$; it decreases as $|\CV_i|$ is increased up to $|\CV_i| = L-t$, but then remains fixed when $L-t \le |\CV_i| \le L$. It follows that, for a given active AP $h_i$, there are only $z_i=\min(L-\Sfl(i),L-t)$ different rate values that the users in a feasible set can have. So, if $L-\Sfl(i)>L-t$, one does not need to consider the feasible sets $\CV_i$ with $L-t \le |\CV_i| < L-\Sfl(i)$, as the users' individual rates won't change and the rate vector will be dominated by the rate vector using all the $L-\Sfl(i)$ users in codeword formation (similar discussion appeared also in Remark~\ref{rem:complexity_remark}). As a result, for a given active AP $h_i$, it is sufficient to search for the largest weighted sum-rate among the $z_i$ different rate vectors, where only the users with the largest queue backlogs in $\CC_i$ are chosen in the codeword formation. This is accomplished by sorting the users in $\CC_i$ by their queue backlogs, from the largest to the smallest. Then, we create feasible sets $\CV_{i,g}$, $g\in[z_i]$, such that for $g < z_i$, $\CV_{i,g}$ consists of the first $g$ users among the sorted ones, and for $g=z_i$, $\CV_{i,g}$ consists of all the $L-\Sfl(i)$ users in $\CC_i$.
% , $\CV_{i,2}$ consists of the first two users; the third set $\CV_{i,3}$ consists of the first three users; and so on, with the second-to-last set $\CV_{i,z_i-1}$ consisting of the first $z_i-1$ users and the last set $\CV_{i,z_i}$ of all the $L-\Sfl(i)$ users. %Notice that there are $L-\Sfl(i)$ users since there are $\Sfl(i)$  number of cache profiles where $U_i^l$ is empty. 
By constructing the corresponding rate vector for each $\CV_{i,g}$
%((Then, we create feasible sets $\CV_{i,g}$, $g\in[z_i]$, such that for $g < z_i$,  $\CV_{i,g}$ consists of the first $g$ users among the sorted ones and for $g=z_i$, $\CV_{i,g}$ consists of all the $L-\Sfl(i)$ users. We construct the corresponding rate vector for each $\CV_{i,g}$ and by ...))
%(instantaneous rate of users is calculated by~\eqref{eq:numberofslots} and~\eqref{inst-rate-expression}).
%where all elements are zero, except for the calculated instantaneous rate at the indices of the users in the feasible set. 
and solving the weighted sum-rate maximization over these $z_i$ rate vectors, we can find the rate vector $\Br_{(i)}$ achieving the solution for an active AP $h_i$. Let  $w_{(i)}$ denote the achieved weighted sum-rate for $\Br_{(i)}$. Then, the instantaneous rate vector for the activation pattern $\Bp_j$ is given by $\Br(j,1)=\sum_{\Bp_j[i] = 1}\Br_{(i)} $ and the corresponding weighted sum-rate by $w_j=\sum_{\Bp_j[i] = 1}w_{(i)}$.
This is conveniently illustrated by the following example. 
 \begin{exmp}
\label{ex:dynmalgorithm}
    Let $L=5$, $t=2$, and w.l.o.g., for a given activation pattern $\Bp_j$ and active AP $h_i$, let $\Sfl(i)=1$, so there are $4$ cache profiles where $U_i^l$ is not empty. Assume also that $\CC_i=\lbrace u_1,u_2,u_3,u_4 \rbrace$, where $Q_1 > Q_2 > Q_3 > Q_4$. %where users $u_1, u_2, u_3, u_4,$ each assigned to a different cache profile, can be served together by the given active AP and have the largest queue backlogs in their respective cache groups. 
    Then, for $h_i$, there are $z_i=3$ constructed rate vectors. The feasible sets, the corresponding rate vectors, and the resulting weighted sum-rates are presented in Table~\ref{tab:wsrAPs}. Notice from Table~\ref{tab:wsrAPs} that since $(1,1,1,0,...,0)$ is dominated by $(1,1,1,1,0,...,0)$, we can exclude it from the solution of the weighted sum-rate maximization.
    %for $\CV_{i,1}=\lbrace u_1 \rbrace$%is served
    %, we get $(\frac{5}{3},0,0,...,0)$, for $\CV_{i,2}= \lbrace u_1, u_2 \rbrace $% are served
    %, we get $(\frac{10}{9},\frac{10}{9},0,0,...,0)$, and for $\CV_{i,3}=\lbrace u_1, u_2, u_3, u_4 \rbrace$ we get $(1,1,1,1,0,0,...,0)$. Notice that if only $u_1,u_2,u_3$ are chosen, we get $(1,1,1,0,0,...,0)$; however this rate vector is dominated by $(1,1,1,1,0,0,...,0)$ so we disregard it. Then, wsrm is solved over these three rate vectors, i.e., we find the rate vector that achieves the maximum between $ (\frac{5}{3}Q_1,\frac{10}{9}(Q_1+Q_2),(Q_1+Q_2+Q_3+Q_4))$. %And after solving the wsrm %similarly 
%for each active helper independently, we sum the constructed rate vectors to attain the instantaneous rate vector corresponding to the activation pattern, and sum the achieved weighted sum rates of
\end{exmp}
\begin{table}
    \centering
    
  %  \vspace{5pt}
    \begin{tabular}{c|c|c}
         \textbf{Feasible set} & \textbf{Rate vector} & \textbf{Weighted sum-rate}  \\
         \hline
         \hline
         $\CV_{i,1}=\lbrace u_1 \rbrace$ & $(\frac{5}{3},0,...,0)$ & $\frac{5}{3}Q_1$   \\
         \hline
         $\CV_{i,2}= \lbrace u_1, u_2 \rbrace $ & $(\frac{10}{9},\frac{10}{9},0,...,0)$ & $\frac{10}{9}(Q_1+Q_2)$ \\
         \hline
         $\CV_{i,3}=\lbrace u_1, u_2, u_3, u_4 \rbrace$ & $(1,1,1,1,0,...,0)$ & $Q_1+Q_2+Q_3+Q_4$    \\
         \hline
         $\lbrace u_1,u_2,u_3\rbrace$ & $(1,1,1,0,...,0)$ & $Q_1+Q_2+Q_3$  
    \end{tabular}
    \caption{The feasible sets, the corresponding rate vectors, and the resulting weighted sum-rates for the active AP $h_i$ in Example~\ref{ex:dynmalgorithm}.}
   
   \label{tab:wsrAPs}
\end{table}

Steps 1 and 2 provide an instantaneous rate vector for each activation pattern and its corresponding weighted sum-rate. Then, to find the solution to~\eqref{instantaneous-wsrm}, we simply determine the activation pattern $j^*$ that maximizes the weighted sum-rate. So, the solution to~\eqref{instantaneous-wsrm} is given by $\Br(j^*,1)$ where $j^*=\arg\max_{j \in [2^H-1]}{w_j}$.
%After calculating $w_j$ for each activation pattern $p_j$, $\Br(j^*,1)$ gives the solution to~\eqref{instantaneous-wsrm} where $j^*=\arg\max_{j \in [2^H-1]}{w_j}$. 
%
This solution reduces the number of scheduling decisions from~\eqref{eq:complexity_calc_opt} to\footnote{Notice that for a given activation pattern $\Bp_j$ and active AP $h_i$, if $z_i=1$, there is a unique resulting weighted sum-rate for $h_i$. Hence, $h_i$ can be excluded from the computation in~\eqref{eq:dynamic_complexity} for $\Bp_j$. Furthermore, if for a given $\Bp_j$ all active APs $h_i$ satisfy $z_i=1$, the number of scheduling decisions corresponding to $\Bp_j$ reduces to one.} 
%
%
%\textcolor{red}{In the second step, we find} %\eqref{instantaneous-wsrm} is replaced by 
%\begin{equation} 
%\hat{\BR}_t = \arg\max_{\hat{\Br} \in \{\hat{\Br}_j\}} \; \hat{\Br}^{\sf T} \BQ_{t}, \label{instantaneous-wsrm-superuser}
%\end{equation}
%where $\{\hat{\Br}_j\}$ is the set of instantaneous rate vectors corresponding to the $2^H-1$ activation patterns. \textcolor{red}{Notice that in the first step given an activation pattern each active helper}
%
\begin{equation}
    \label{eq:dynamic_complexity}
    \sum_{j \in [2^H-1]}\sum_{\substack{i \in [H],\\\Bp_j[i] = 1}}
    \min (L-\Sfl(i),L-t),
\end{equation}
indicating a massive decrease in the number of considered scheduling decisions. For instance, notice that the term inside the summation over the activation patterns $j \in [2^H-1]$ is linear in $H$ for fixed $L$, instead of exponential as in~\eqref{eq:complexity_calc_opt}. 
As an example, consider the same parameters as in the paragraph following equation~\eqref{eq:complexity_calc_opt}: a network with $H=4$ APs and $L=4$, where $2$ users from each cache profile and a total of $8$ users can be served by each AP simultaneously. Assume also $t=1$. Then, for the activation pattern $p_j=(1,1,1,1)$, by using the reduced-complexity dynamic solution, the number of scheduling decisions is reduced from $O(10^7)$ to just $12$.

\subsection{The Virtual Queue Heuristic}
\label{section:vq_heur}

%Here 
We propose a heuristic for the DPP dynamic scheduler, suitable for very large networks where even the %computation of the $2^H-1$ instantaneous rate vectors of the super-user reduced network is 
reduced-complexity solution in Section~\ref{sec:new_dynamic} is impractical. 
For these cases, we aim at 
%When  or even the %reduced \textcolor{red}{solution in %\eqref{instantaneous-wsrm-superuser} Section~\ref{sec:new_dynamic}} is too complex, we can 
obtaining a suboptimal but scalable to very large networks DPP dynamic scheduler by replacing the instantaneous weighted sum-rate maximization step in~\eqref{instantaneous-wsrm} with a heuristic choice of the AP activation pattern and corresponding CC codewords. 
%
%\centerline{\textcolor{red}{\bf END OF REVISION}}
%
%
%\textcolor{blue}{[.... KAGAN: since we have already said that the solution of the weighted sum-rate maximizaiton is the complexity bottleneck, there is no need to repeat later. In fact, you have **first**
%to explain the meaning of the algorithm, i.e., that we favor to serve users with a large virtual queue, 
%and we do so by finding a feasible activation pattern that gives priority to these users, 
%and find the corresponding CC codeword and therefore the rate vector. This is done by solrting the users, scanning the network by looking at an activation pattern of the APs, etc etc ... QUALITATIVE EXPLANAITON. 
%Then ... you define a notation, and give the pseudo-code.]}
%
%\textcolor{red}{[OLD TEXT]}
%To address this issue, we propose a new low-complexity method to
%In the heuristic, 
The idea is to favor serving users with large queue backlogs. The reasoning is that the weighted sum-rate in~\eqref{instantaneous-wsrm} increases when higher rates are assigned to such users. So, instead of solving~\eqref{instantaneous-wsrm} or the reduced-complexity dynamic solution by searching over all activation patterns, we find a feasible activation pattern that prioritizes the users with large queue backlogs by assigning them to APs. %, which then 
The corresponding codewords and the rate vector are then determined accordingly.

As a general description of the proposed heuristic, at each scheduling slot, we sort users by their virtual queue backlogs from largest to smallest. Then, the users are assigned to APs one by one. %Notice that, in this way, the activation pattern and the corresponding codeword formation are found 
% We find a suitable rate vector at every scheduling slot by sorting the users from the largest virtual queue length to the smallest and greedily assigning APs to users one by one. The reasoning behind this is that the expression in the maximization in (\ref{wrsm_eq}), i.e., the weighted sum rate, increases when larger rates are assigned to users with larger virtual queue lengths. 
%Now, continuing with the general description of the algorithm, at every iteration, the algorithm first sorts the users according to their virtual queue lengths and 
To assign each user to an AP, we search for suitable APs: %to be assigned to users. 
If there is a unique active AP that can %potentially 
serve a user %, the AP is assigned to the user, provided that 
and the user is not interfered with by another active AP, we check the weighted sum-rate in~\eqref{instantaneous-wsrm}, and assign the user to the AP only if the resulting weighted sum-rate increases. 
If there is no active AP that can serve the user, an inactive AP is randomly chosen among those that can serve the user (if any), provided that the user isn't interfered with by any currently active AP. In both cases, it is ensured that each active AP serves at most one user assigned to each cache profile, and that interfering inactive APs remain inactive in the same scheduling slot. 
%It is easy to see that for large enough $T$, all the users will be served since the unserved users will have the largest queues in the long run and will be chosen to be served eventually. 

%In particular, the steps of the heuristic at each scheduling slot is given by the pseudo-code in Algorithm~\ref{alg:main_vq}.
%here we propose the scheme summarized
The steps of the proposed Virtual Queue Heuristic at each scheduling slot are given by the pseudocode in Algorithm~\ref{alg:main_vq} 
%
%\textcolor{blue}{[ .... KAGAN: it is very important the following: 
%\begin{enumerate}
%    \item You must make very clear that the only ``memory'' between successive scheduling slots 
%    $t = 1,2,3,\ldots$ in this algorithm is represented by the virtual queue backlogs $\BQ_t$ ... in your description it is not clear when you say ``iteration'', or when you ``initialize'' quantities, if this refer to what it is done at each $t$, that is, you initialize all this stuff apart from the queues at each $t$, or if certain things are initialized at the beginning for $t = 0$ and then do not reset. 
%    YOU NEED TO DISTINGUISH BETWEEN TIME SLOTS, AND ``ITERATIONS'' OF THE ALGORITHM TO SELECT THE SCHEDULIGN DECISION AT EACH TIME SLOT!!!
%    \item The only initialization at the beginning of time is $\BQ_0$. Then, there is not point in giving a number of steps $T$ .. the algorithm repeats for each $t$ forever ... over an infinite time horizon. 
%    \item You need to say that at each $t$, this and this and this things are initialized, and then the APs are activated and the CC codewords (and corresponding rates) are selected via iterations over the network (not over time).
%    \item You need to replace line 25 of the pseudocode with
%    ``compute the virtual arrival vector using \eqref{virtual_arrival}''
%    \item You need to replace line 26 of the pseudocode with
%    ``update the virtual queue backlog using \eqref{queue_eq}'' where you have defined $\BR_t$ in the pseudocode consistently with the notation intorduced in the general DPP (see before).    
%\end{enumerate}
%]}
%
where we use the following notations:
1) $t$ is the scheduling slot index; 
2) $\CH_{\mathrm{all}}$ denotes the set of all APs; %and 
3) $\CH$ denotes the set of candidate APs to be activated;
%At the beginning of every iteration, we set $\CH=\CH_{\mathrm{all}}$;
4) $\CH_{\mathrm{on}}$ denotes the set of active APs; 
%At the beginning of every iteration, all APs are  assumed inactive, i.e., we start with the initialization $\CH_{\mathrm{on}} = \varnothing$;
%, and $T$ is the total number of slots the algorithm runs (this could be infinite); 4) $\BQ_t$ denotes the virtual queue backlog vector at scheduling slot $t$;
5) $w_{i}^{temp}$ denotes the temporary weighted sum-rate of the AP $h_i$;
6) $\Bm_i$, $i \in [H]$, is a vector of length $L$ that expresses the ``local cache population'' of AP $h_i$, i.e., %it is initially set to all-zero. Then,  
for all $i \in [H]$, 
$\Bm_i[l] = k$ only if AP $h_i$ is serving $u_k$ and $\SfL(u_k) = l$ (otherwise, it is zero); and $\lbrace\Bm_i\rbrace_s$ denotes the set of users %indices 
included in $\Bm_i$. We have also used some auxiliary functions:
1) $\textsc{Order}(\Bu)$ sorts the set of users $\{u_k\}$ according to $\BQ_t[k]$ values, 
in descending order;
%from the largest to the smallest;
%from largest $Q_k$ to smallest, $Q_k$ will be defined later;
2) $\textsc{Inter}(u_k)$ gives the set of active APs in $\CH_{\mathrm{on}}$ which are within radius $r_{\mathrm{inter}}$ of user $u_k$; %2) $\textsc{Total}(u_k)$ gives the number of active APs in $\CH_{\mathrm{on}}$ which are within radius $r_{\mathrm{inter}}$ of user $u_k$. 
3) $\textsc{Trans}(u_k)$ gives the set of APs in $\CH$ which are within radius $r_{\mathrm{trans}}$ of user $u_k$; and 
%6) $\BQ$ denotes the virtual queue backlog vector at scheduling slot $t$. 
4) $\textsc{Wsr}(\Bm_i)$ gives the weighted sum-rate of AP $h_i$, assuming the users in $\Bm_i$ are assigned to $h_i$.

At the beginning of every scheduling slot $t$, we set $\CH=\CH_{\mathrm{all}}$, $\CH_{\mathrm{on}} = \varnothing$, and $\Bm_i$ to all-zero vector for every $i \in [H]$. (Line 1 of Algorithm~\ref{alg:main_vq}). The algorithm runs indefinitely for each scheduling slot $t$, where users are assigned to APs and the resulting instantaneous rate vector $\Br$ is calculated via iterations over the network.  %In addition, at the beginning of Algorithm~\ref{alg:main_vq}, i.e., at $t=0$, we set $\BQ=\bf{0}$.

\begin{algorithm}[h]
%\small
\caption{Virtual Queue Heuristic}
\label{alg:main_vq}
\begin{algorithmic}[1]
\State $\textsc{Initialize}$
%\While{$t<T$}
\State $\textsc{Order}(\Bu)$
\ForAll{$u_k \in \Bu$}
 \State $\CI_k \gets \textsc{Inter}(u_k)$
 \If{$|\CI_k| == 1$}
  \State $h_i \gets \CI_k[1]$ 
   %\If{$\textsc{Total}(u_k)=1$}
    %\State $h_i \gets \textsc{Find}(u_k)$
  \If{$|\Sfx(h_i)-\Sfx(u_k)| \le r_{\mathrm{trans}}$, $\Bm_i[\SfL(u_k)] = 0$}
   \State $\Bm_i[\SfL(u_k)] \gets k$ 
   \If{$\textsc{Wsr}(\Bm_i) < w^{temp}_i$}
   \State $\Bm_i[\SfL(u_k)] \gets 0$
   \Else 
   \ForAll{$h_{i'} \in \CH$, $h_{i'}\neq{h_i}$}
    \If{$|\Sfx(h_{i'})-\Sfx(u_k)| \le r_{\mathrm{inter}}$}
     \State Remove $h_{i'}$ from $\CH$
    \EndIf
   \EndFor
   \EndIf
  \EndIf
 \EndIf
  \If{$|\CI_k| == 0$}
   \State $\CT_k \gets \textsc{Trans}(u_k)$
   \If{$|\CT_k|>0$}
    \State $h_i \gets \CT_k[rng]$
    \State $\Bm_i[\SfL(u_k)] \gets k$
    \State  $w_i^{temp} \gets \textsc{Wsr}(\Bm_i)$  
    \State Add $h_i$ to $\CH_{\mathrm{on}}$
    \ForAll{$h_{i'} \in \CH$, $h_{i'}\neq{h_i}$}
        \If{$|\Sfx(h_{i'})-\Sfx(u_k)| \le r_{\mathrm{inter}}$}
          \State Remove $h_{i'}$ from $\CH$
        \EndIf
    \EndFor  
   \EndIf
  \EndIf
\EndFor
\ForAll{$h_i \in \CH$}
  \ForAll{$l \in [L]$, $\Bm_i[l] \neq 0$}
   \State $\Br[\Bm_i[l]] \gets r(\lbrace\Bm_i\rbrace_s)$
   %r_{k(i)}$
  \EndFor
\EndFor
\State Compute the virtual arrival vector using \eqref{virtual_arrival}
\State Update the virtual queue backlog using \eqref{queue_eq}
\If{$t>1$}
  \State $\bar{\Br}[k] \gets (\bar{\Br}[k]*(t-1)+\Br[k])/t$
  \Else
  \State $\bar{\Br}[k] \gets \Br[k]$
\EndIf
\State $t \gets t+1$
%\State $\CH \gets \CH_{\mathrm{all}}$
%\EndWhile
\end{algorithmic}
\end{algorithm}

%%%%%%%%%%%%%%%%%%%%%%%%%%%%%%%%%%%%%%%%%%%%%%%%%%%%%%%%%%%%%%
%%%%%%%%%%%%%%%%%%%%%%%%%%%%%%%%%%%%%%%%%%%%%%%%%%%%%%%%%%%%%%
\section{Numerical Results}
\label{section:sim_results}

%\todo[inline]{Please provide your explanation for each figure: what are the key takeaways and the explanation of various behaviors we see there.}

%\input{SimResults.tex}

%First, we show, by example, how users' goodputs adapt to changes in a dynamic network.

%First, we simulate a dynamic network example to show how users' goodputs adapt to changes in the network.
%Then, we provide simulations to compare the performance of the optimum solution, i.e., the solution of~\eqref{eq:optimization_problem_main} or equivalently the goodput vector achieved by~\eqref{queue_eq}-\eqref{instantaneous-wsrm}, %super-user (Section~\ref{section:super-user}), 
%with the Virtual Queue Heuristic (Section~\ref{section:vq_heur}) solution and two other baseline schemes.

First, we present simulations comparing the performance of the optimum solution, i.e., the solution to~\eqref{eq:optimization_problem_main}, or equivalently, the goodput vector achieved by~\eqref{queue_eq}–\eqref{instantaneous-wsrm}, with that of: 1)  the Virtual Queue Heuristic (Section~\ref{section:vq_heur}); 2) a conventional uncoded caching scheme (see Remark~\ref{uncoded-caching-rem}), and 3) two additional baseline schemes described below. Then, we simulate a dynamic network scenario to illustrate how users’ goodputs adapt to changes in the network. 
%((This sentence should be written better.)) % and the RGA algorithm of~\cite{akcay_gc}. As a quick explanation, the RGA heuristic categorizes the users according to the number of helpers within their interference radius and assigns them to helpers randomly in a greedy fashion. 
%For comparison, we also simulate two methods based on virtual queue theory. %The difference between the two methods lies in how the helpers are activated; however, in both methods, collisions are avoided by not activating the neighboring helpers at the same slot. %the reuse with multi-round delivery approach of~\cite{mozhgan}. %with a frequency reuse factor of three

%rate at which chunks are made available to the user as the
%goodput of the user.

\textbf{Streaming rates:} We assume that each AP has the same downlink multicast channel capacity, denoted by $C$. Recall that the goodput of a user is defined as the time-averaged rate (normalized rate per unit time) at which video chunks are made available to the user. Thus, for a given goodput $\bar{r}$, the corresponding physical-layer rate at which a user receives video chunks is $C\bar{r}$ bit/s. In all the subsequent figures, each reported goodput value can be multiplied by the AP capacity $C$ to obtain the actual %physical rate.
(physical) streaming rate. 
%with the RGA heuristic, the users are categorized according to the number of helpers within their interference radius and assigned to helpers randomly in a greedy fashion. On the other hand, 

\textbf{Baseline schemes:} In addition to conventional (uncoded) caching, we use two other baseline schemes for comparison. 

The first baseline, called \emph{reuse with queue update}, uses standard channel reuse to avoid interference. 
%In the multi-round delivery method, the users assigned to the same AP are served sequentially in groups, where the groups are selected such that the users in the same group have different cache profiles. 
%For a given helper set where there is no interference among the helpers, i.e., when 
%
%Channel reuse consists of avoiding mutual interference of neighboring APs by allocating them to 
%orthogonal channels so that they do not collide. 
A channel reuse scheme with reuse factor $m$ avoids mutual interference between neighboring APs by allocating them to $m$ orthogonal channels so they do not collide. 
Each sub-channel is activated in round-robin (in time) or parallel on $1/m$ of the available bandwidth (in frequency). In our case, this corresponds to activating the non-neighboring APs in the same sub-channel every $m$-th slot.

The second baseline, called \emph{CSMA-inspired queue update}, as its name suggests, is inspired by idealized CSMA models~\cite {csma1,csma2}. %, where nodes that wish to transmit, if the channel is sensed ``free'', 
%wait an exponentially distributed back-off time.
With this baseline, each AP is assigned a timer from an exponential distribution at the beginning of each scheduling slot. 
Depending on the timer values, APs are activated one by one, %provided that the neighboring helpers of activated helpers remain inactive 
provided that they don’t create interference
with users that are already being served by other active APs, obeying
the ready-to-send/clear-to-send distributed coordination function. % For instance, for a network with $3$ helpers, let the assigned timers of the helpers $h_1$, $h_2$ and $h_3$ be $a$, $b$ and $c$, respectively. Assume $a<b<c$. First, $h_1$ is activated, and then $h_2$ is activated if $h_2$ is not a neighbor of $h_1$; otherwise, $h_2$ remains inactive, then $h_3$ is activated if it is not a neighbor of $h_1$ and also not a neighbor of $h_2$ in case $h_2$ is activated. In both methods, given an activation pattern at a slot, each active helper updates the virtual queues~\eqref{queue_eq}~-~\eqref{wrsm_eq} locally for only users within its transmission radius.
In both baseline schemes, interference is avoided. Therefore, 
the wsrm problem~\eqref{instantaneous-wsrm} restricted to the allowed AP activation patterns decouples into independent maximizations, one for each active AP.  %In other words, the baseline schemes solve the interference problem using a standard collision-avoidance technique (reuse, or CSMA), thereby reducing the weighted sum-rate maximization to a set of mutually disjoint smaller problems, one for each (active) AP. 
In other words, by using a standard collision-avoidance technique (reuse or CSMA), the baseline schemes reduce the weighted sum-rate maximization in~\eqref{instantaneous-wsrm} to a set of mutually disjoint smaller problems that can be solved with low complexity.

\textbf{Simulation setup:} We assume that $H$ APs are located at the center of hexagons on a finite hexagonal grid. Each hexagon has a radius of~$1$ (normalized length unit), we set $r_{\mathrm{trans}} = 1$ and $r_{\mathrm{inter}}  =1.2$. Notice that since the AP locations form a hexagonal grid, the minimum spatial reuse factor required to guarantee collision-free transmissions is $m = 3$.

\begin{figure}[t]
    \centering
    \includegraphics[width=0.55\linewidth]{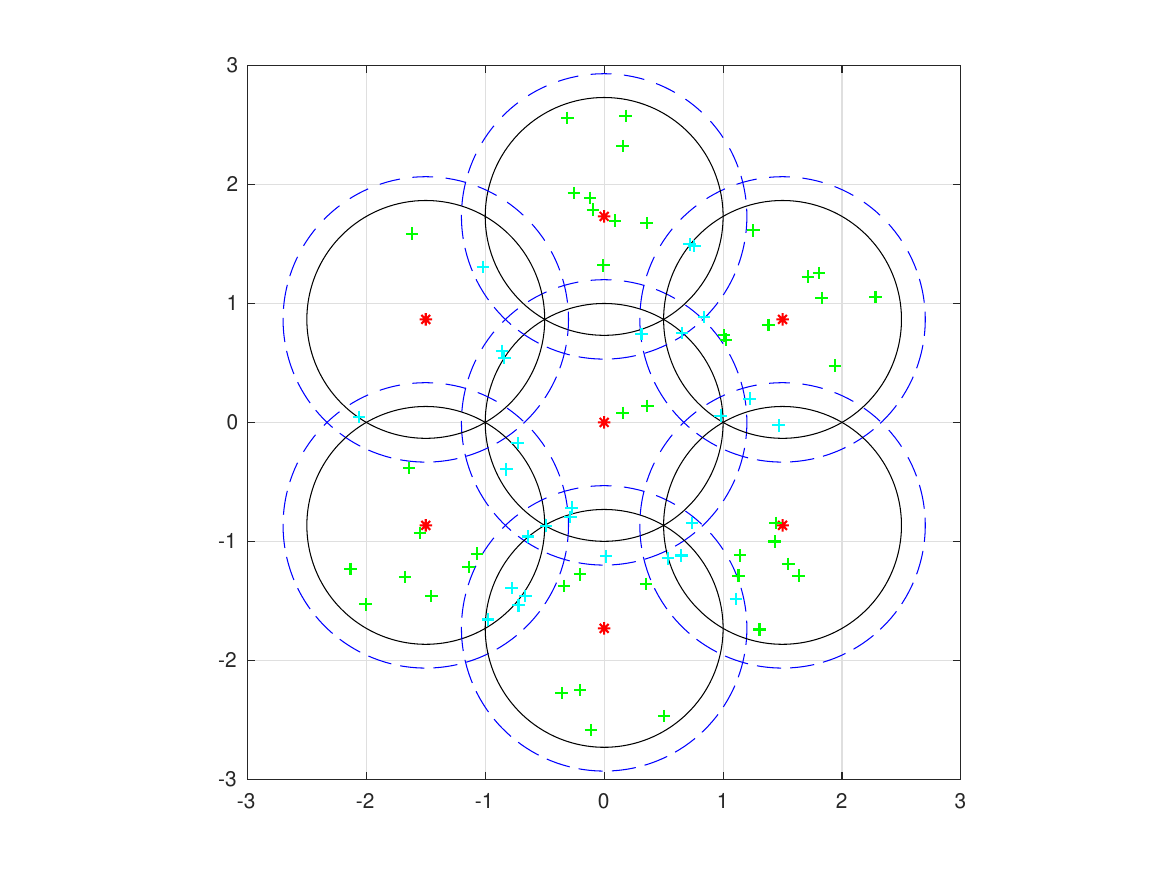}
    \caption{Example network with $H=7$, $K=70$. Each green user is within the transmission radius of exactly $1$ AP and outside the interference radii of other APs. Each blue user is within multiple AP transmission/interference radii. }
    \label{fig:example_hexagonal}
\end{figure}

\begin{table}
    \centering
  %  \vspace{5pt}
    \begin{tabular}{c||c|c|c}
         Users & Green & Blue & All  \\
         \hline
         \hline
         Optimum Solution & 0.27 & 0.22 & 0.14  \\
         \hline
         Virtual Queue Heuristic & 0.27 & 0.20 & 0.13 \\
         \hline
         CSMA-Inspired Queue Update & 0.27 & 0.17 & 0.10   \\
         \hline
         Reuse with Queue Update & 0.09 & 0.14 & 0.07 
    \end{tabular}
    \caption{Goodput geometric mean comparison of the users in Figure~\ref{fig:example_hexagonal} for PF, with $L=5$, $\gamma=0.2$; if only blue, only green, or all users are served.}
   \label{tab:differentregion}
\end{table}

In Figure~\ref{fig:example_hexagonal} we plot a hexagonal grid that consists of $H=7$ APs and $K=70$ users. Each user is randomly placed within the transmission area of APs. %and randomly assigned to a cache profile $l \in [L]$. %Table~\ref{tab:differentregion} presents the geometric mean of the user goodputs of the network in Figure~\ref{fig:example_hexagonal} for different user subsets.
The geometric means of the user goodputs for PF for the different user subsets are presented in Table~\ref{tab:differentregion}, where each user is randomly assigned to a cache profile $\Phi_l$, $l \in [L=5]$. 

We can see from Table~\ref{tab:differentregion} that when only green users are served, i.e., users located within the transmission radius of a single AP and outside the interference radii of all other APs, the geometric means of the optimum solution and CSMA-Inspired Queue Update are identical. This result is expected, since in this scenario every AP can be activated at each scheduling slot, and regardless of which users an AP serves, those users won't be interfered with by any other active AP. In both methods, each AP independently maximizes its weighted sum-rate, which is optimal. %In the Virtual Queue Heuristic, each AP is also activated at every scheduling slot. %However, since for a given AP, the weighted sum rates of all user multicast groups aren't checked 
In the Virtual Queue Heuristic, all APs are also activated at each scheduling slot; however, because the weighted sum rates of all multicast groups are not exhaustively evaluated (Recall from Section~\ref{section:vq_heur} that the weighted sum-rate is only checked when adding the next user), %the Virtual Queue Heuristic is not necessarily optimal. 
optimality is not guaranteed in general.
Nevertheless, we can see from Table~\ref{tab:differentregion} that %for the network in Figure~\ref{fig:example_hexagonal}, when only green users are served, the Virtual Queue Heuristic also achieves the optimum. 
for the considered network realization, the Virtual Queue Heuristic also attains the optimal performance when only green users are served.

\begin{figure}[t]
    \centering
    \resizebox{0.7\columnwidth}{!}{%
    
    \begin{tikzpicture}

    \begin{axis}
    [
    % put axis lines at left and bottom
    axis lines = left,
    % control axis labels
    xlabel = \smaller {L},
    ylabel = \smaller {Goodput Geometric Mean},
    xmin=0, xmax=40,
    ymin=0, ymax=0.08,
    ylabel near ticks,
    % control legend position
    legend pos = south east,
    % control size of tick marks (10,20,30,etc)
    tick label style={font=\smaller},
    % control major grids
    grid=both,
    major grid style={line width=.2pt,draw=gray!30},
    ytick={0,0.01,0.02,0.03,0.04,0.05,0.06,0.07,0.08},
    % control minor grids
    %grid style={line width=.1pt, draw=gray!10},
    %minor tick num=5,
    ]
    
    % \addplot[black]
    % table[y=Opt-L1-Y,x=Opt-L1-X]{Figs/CDF_data_L.tex};
    % \addlegendentry{\smaller $L=1$, Analytical}
    % \addplot[black,dashed]
    % table[y=Super-L1-Y,x=Super-L1-X]{Figs/CDF_data_L.tex};
    % \addlegendentry{\smaller $L=1$, Super-user}
    % \addplot[black!50]
    % table[y=Heur-L1-Y,x=Heur-L1-X]{Figs/CDF_data_L.tex};
    % \addlegendentry{\smaller $L=1$, RGA}
    
    % \addplot[green]
    % table[y=Opt-L5-Y,x=Opt-L5-X]{Figs/CDF_data_L.tex};
    % \addlegendentry{\smaller $L=5$, Analytical}
    % \addplot[green,dashed]
    % table[y=Super-L5-Y,x=Super-L5-X]{Figs/CDF_data_L.tex};
    % \addlegendentry{\smaller $L=5$, Super-user}
    % \addplot[green!50]
    % table[y=Heur-L5-Y,x=Heur-L5-X]{Figs/CDF_data_L.tex};
    % \addlegendentry{\smaller $L=5$, RGA}
    \addplot[black,dashed,mark=*]
    %coordinates{(1,0.047113392248165)(10,0.060560650051093)(20,
    %0.165211221097960)
    %0.064107488967244)(30,
    %0.166688985691624)
    %0.065348989253869)(40,
    %0.169887458246690)
    %0.065991198848787)};  first of the previuos only using max cc vectors
    coordinates{(1,0.047563186575870)(10,0.063565331812875)(20,0.067328346175086)(30,0.069139818820066)(40,0.070559340413863)};
    \addlegendentry{\smaller Optimum Solution}
    \addplot[blue,dashed,mark=*]
    %table[y=Opt-L1-Y,x=Opt-L1-X]{Figs/CDF_data_L.tex};
    %coordinates{(1,0.039111060849618)(10,0.050034650312476)(20,
    %0.143326727009363)
    %0.051605005522196)(30,
    %0.144812846747907)
    %0.052712928395644)(40,
    %0.148654666592693)
    %0.052865882567710)};
    %table[y=1,x=2];  
    coordinates{(1,0.044329384894313)(10,0.054412117771441)(20,0.056908497001359)(30,0.057812948353130)(40,0.058174249824428)};
    \addlegendentry{\smaller Virtual Queue Heuristic}
   % \addplot[red,dashed,mark=*]
   % table[y=Opt-L5-Y,x=Opt-L5-X]{Figs/CDF_data_L.tex};
   %coordinates{(1,0.0948629508569942)(5,0.136016761501415)(10,
   %0.154405932706313
   %0.151786467250979)(15,
   %0.159354933135196
   %0.160083952235497)(20,
   %0.163571935511174
   %0.162648826627326)};
    %\addlegendentry{\smaller Super-User}
    \addplot[red,dashed,mark=*]
   % table[y=Opt-L10-Y,x=Opt-L10-X]{Figs/CDF_data_L.tex};
   %coordinates{(1,0.080823137585595)(5,0.118390277798257)(10,
   %0.129058973970177
   %0.125713970994251)(15,
   %0.131152879192204
   %0.130417568949779)(20,
   %0.134195027504770
   %0.133624509805899)};
    %\addlegendentry{\smaller RGA}
    %\addplot[orange,dashed,mark=*]
   % table[y=Opt-L15-Y,x=Opt-L15-X]{Figs/CDF_data_L.tex};
  % coordinates{(1,0.041325311987972)(10,0.037232539124400)(20,
   %0.0702163636806340
  % 0.037986062126483)(30,
   %0.0738924658999816
  % 0.039239771695648)(40,
   %0.0776596367561371
  % 0.040084221055906)};
  coordinates{(1,0.041583628047792)(10,0.037722173449604)(20,0.038528655492272)(30,0.039426642759481)(40,0.040473389291529)};
    \addlegendentry{\smaller CSMA-Inspired Queue Update}
    \addplot[orange,dashed,mark=*]
    %coordinates{(1,0.018335457401870)(10,0.029636013032292)(20,0.034409353734709)(30,0.036935579024923)(40,0.038505324718276)};
    coordinates{(1,0.018420294686109)(10,0.029673995111468)(20,0.034538381283504)(30,0.037035721990944)(40,0.038844356778306)};
    \addlegendentry{\smaller Reuse with Queue Update}
    
    %\addplot[gray]
    %table[y=Opt-L20-Y,x=Opt-L20-X]{Figs/CDF_data_L.tex};
    %\addlegendentry{\smaller $L=20$}

    \end{axis}

    \end{tikzpicture}
    }
    \caption{Goodput geometric mean for different values of $L$ for PF, with $H=10$, $K_{avg}=200$, $\gamma=0.1$.}
    \label{fig:G_changeL}
\end{figure}
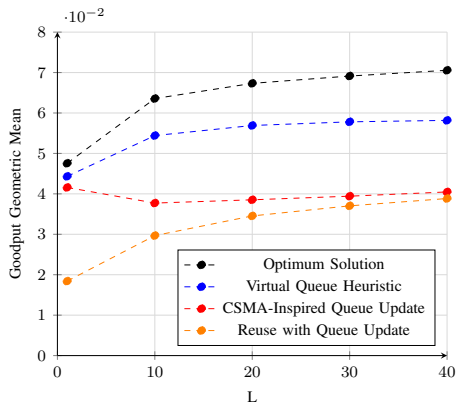

In contrast, Reuse with Queue Update assigns a priori distinct reuse numbers to neighboring APs, %so that only APs with the same reuse number are activated simultaneously. 
allowing only APs with the same reuse number to be active simultaneously. %However, there would be no interference even if all APs were active, so assigning different reuse numbers is unnecessary. Actually, since each AP is activated only one at each $ 3$rd slot, the geometric mean of the reuse method is $1/3$ of the optimal geometric mean.  %This shows that, even 
%Goodput geometric mean comparison of the users in Figure~\ref{fig:example_hexagonal} for pf, with $L=5$, $\gamma=0.2$; if only blue, only green, or all users are served.
In the absence of interference, such reuse constraints are unnecessary. Since each AP is activated only once every three slots, the resulting geometric mean goodput is approximately one-third of the optimal value.

%In contrast, the Reuse with Queue Update scheme assigns distinct reuse indices to neighboring APs, allowing only APs with the same reuse index to be active simultaneously. In the absence of interference, such reuse constraints are unnecessary. Since each AP is activated only once every three slots, the resulting geometric mean goodput is approximately one-third of the optimal value.

%When only green users are served, i.e., users located within the transmission radius of a single AP and outside the interference radii of all other APs, the geometric means achieved by the optimal solution and the CSMA-inspired queue update are identical. This result is expected, since in this scenario all APs can be activated in every scheduling slot without causing interference. Consequently, each AP independently maximizes its weighted sum rate, which is globally optimal. In the Virtual Queue Heuristic, all APs are also activated at each scheduling slot; however, because the weighted sum rates of all multicast groups are not exhaustively evaluated (see Section~\ref{section:vq_heur}), optimality is not guaranteed in general. Nevertheless, for the considered network realization, the Virtual Queue Heuristic also attains the optimal performance when only green users are served.

When only blue users are served, i.e., users located within multiple AP transmission/interference radii, both the Virtual Queue Heuristic and the CSMA-Inspired Queue Update become suboptimal, as shown in Table~\ref{tab:differentregion}. The reason is that when multiple APs are activated simultaneously, interference occurs among some users. The Virtual Queue Heuristic sorts the user queue backlogs and assigns each user to an appropriate AP sequentially. Even though the activation pattern depends on the queue backlogs since users with larger queue backlogs are prioritized, the choice of the activation pattern is still heuristic because the solution is not searched over all activation patterns, as in the optimum solution. This prevents some users from receiving any transmission in a given scheduling slot, even though they might receive transmission in the optimum solution. In CSMA-Inspired Queue Update, the activation order of the APs is random and independent of the queue backlogs, which further degrades performance and leads to worse results than the Virtual Queue Heuristic.

% In this case, simultaneous activation of multiple APs leads to inter-AP interference. The Virtual Queue Heuristic assigns users sequentially based on queue backlogs, but the activation pattern is not optimized over all feasible combinations, unlike the optimal solution. As a result, some users may not be scheduled in a given slot even though they would be served under the optimal policy. In the CSMA-inspired queue update, AP activation is random and independent of queue backlogs, which further degrades performance and leads to worse results than the Virtual Queue Heuristic.

%On the other hand, the geometric mean of the Reuse with Queue Update is closer to the geometric means of the other schemes than in the case where only green users are served. This is because now there will be interference among some users if multiple APs are activated, so interference avoidance by a reuse pattern is more plausible. However, this method remains suboptimal because not every user is in transmission/interference radii of 3 APs. 
The Reuse with Queue Update performs closer to the other methods in this case, since interference avoidance via reuse becomes more relevant when serving blue users. However, it remains suboptimal because the fixed reuse pattern is overly restrictive. For instance, if a user within the transmission/interference radii of only 2 APs is chosen to be served, one of the APs cannot be activated, but another $3$rd AP, neighboring both APs, can still be activated. However, this is not permitted in Reuse with Queue Update, resulting in worse than optimal performance. In fact, Table~\ref{tab:differentregion} shows that it performs the worst. 

%The Reuse with Queue Update scheme performs closer to the other methods in this case, since interference avoidance via reuse becomes more relevant when serving blue users. However, it remains suboptimal because the fixed reuse pattern is overly restrictive. For example, when a user is within the transmission and interference radii of only two APs, it may still be feasible to activate a third neighboring AP without causing interference, which is not permitted by the reuse constraint. Consequently, Reuse with Queue Update yields the lowest geometric mean goodput among all schemes.

Finally, when all users are served, Table~\ref{tab:differentregion} shows that the proposed schemes, i.e., the optimum solution and the Virtual Queue Heuristic, outperform the baseline schemes.

%So, at each scheduling slot not every AP should be activated, the activation pattern and the corresponding users to be served should be chosen carefully.

%the geometric mean of the Reuse with Queue Update is much closer to the other schemes, even though it is still the worst. 

%In VQH, users are sorted according to queue backlog values such that users with high backlog values are prioritized, and since every AP is active and the users' APs can serve are different, at every scheduling slot, the maximum number of users with the highest possible queue backlogs are served, which is also optimal.

%Example network with $H=7$, $K=70$. Each green user is within the transmission radius of only $1$ AP, and outside of the interference radii of other APs. Each blue user is within multiple AP transmission/interference radii.

We also conduct Monte Carlo simulations in which users are distributed according to a homogeneous Poisson point process within the transmission area of the APs. The average number of users in the transmission area, denoted by $K_{avg}$, is determined by the density of the Poisson point process. Each user is randomly assigned to a cache profile $\Phi_l$, $l \in [L]$. We set $\gamma=0.1$, $H=10$, and $K_{avg}=200$, and compare the performance of different schemes in Figures~\ref{fig:G_changeL}-\ref{fig:hf_pf}.
%\textcolor{blue}{

%Finally, we conduct Monte Carlo simulations in which users are distributed according to a homogeneous Poisson point process within the transmission area of the APs.

%It is important to note that 
First, recall from Remark~\ref{uncoded-caching-rem} that the case 
$L=1$ in Figures~\ref{fig:G_changeL} and~\ref{fig:hf_changeL} coincides with the {\em conventional} (i.e., uncoded) 
caching scheme (e.g., ``prefix caching''~\cite{prefixcaching}) and can be applied to each of the considered scheduling/transmission schemes %\textcolor{red}{[add ref]}), 
by assuming that every user caches the same portions of each chunk. In this case, since no multicasting opportunities arise, the server transmits the requested portion of each chunk via standard unicast transmissions.
%In this way, since there are no multicasting opportunities, the server sends the requested part of each chunk via standard unicast transmissions.%}

\input{Figs/New_hf_versus_L}

In Figure~\ref{fig:G_changeL}, we plot the goodput geometric mean versus $L \in \{1,10,20,35,40 \}$ for PF %$\gamma=0.1$, $H=10$, and $K_{avg}=200$ 
for all the schemes. We observe that all the proposed schemes significantly outperform the baseline 
schemes, as expected from the discussion following Table~\ref{tab:differentregion}.  It can also be seen that the geometric means of the optimum solution, the Virtual Queue Heuristic, and the Reuse with Queue Update increase with $L$, reflecting the greater multicast opportunities for larger $L$ values. The important thing to note here is that %the two proposed schemes have a significant gain 
over $L=1$, i.e., uncoded (prefix) caching, as $L$ increases.
there is a significant gain for the two proposed schemes over $L=1$, i.e., uncoded (prefix) caching, as $L$ increases. 

The optimum solution searches for the optimum over all possible multicast possibilities, the Virtual Queue Heuristic finds a desirable activation pattern that gives higher rates to users with larger queue backlogs, and in the Reuse with Queue Update, each AP maximizes its weighted sum rate while creating no interference; so the performance of these schemes can only increase with $L$ since they all benefit from more multicast opportunities. However, as $L$ increases, the rate of increase in the geometric means slows down. This is because $K_{avg}$ is fixed, so the multicast opportunities are limited even if $L$ becomes very large. 

%In contrast, as can be seen from Figure~\ref{fig:G_changeL}, the geometric mean of the CSMA-Inspired Queue Update decreases as $L$ increases from $1$ to $10$. The reason for this is as follows: When an AP is activated, it maximizes its weighted sum rate. For $L=1$, an AP can serve only $1$ user; for $L=10$, an AP can serve up to $10$ users and achieve a higher weighted sum rate than for $L=1$. 
In contrast, as shown in Fig.~\ref{fig:G_changeL}, the geometric mean of the CSMA-Inspired Queue Update decreases as 
$L$ increases from $1$ to $10$. This behavior can be explained as follows. When an AP is activated, it maximizes its weighted sum rate. For $L=1$, an AP can serve only one user, whereas for $L=10$, an AP can serve up to ten users and thus achieve a higher weighted sum rate. However, %in the meantime, an AP serving more users can cause increased network interference, preventing more APs from being activated and other users from being served.
serving a larger number of users at a single AP can increase network interference, which limits the activation of additional APs and prevents other users from being served.

%However, serving a larger number of users at a single AP increases network interference, which limits the activation of additional APs and reduces the number of users that can be served overall.

For the CSMA-Inspired Queue Update, it turns out that the activation order of the APs, being random and independent of the queue backlogs, creates significant interference and offsets the CC advantage for $L=10$. As a result, users with larger queue backlogs cannot be served at sufficiently high rates. Nevertheless, as $L$ continues to increase beyond 10 (e.g, $L=20,30,40$), 
the additional interference caused by serving more users diminishes, particularly in terms of preventing neighboring APs from activating. Consequently, the performance of the CSMA-Inspired Queue Update improves.
%the interference that is created more by serving more users, i.e., in the sense that helpers are prevented from activating, decreases, so the performance of CSMA also increases. 

Furthermore, Figure~\ref{fig:G_changeL} shows that as $L$ increases, %both baseline schemes start converging to each other. 
the performance of both baseline schemes begins to converge. This occurs because, for sufficiently large enough $L$, each activated AP can serve all users within its transmission radius. So in CSMA-Inspired Queue Update, for a dense enough network, neighboring APs can't be activated, and the activation pattern resembles a reuse pattern.  

 %In this regime, the dense network and mutual interference prevent neighboring APs from being simultaneously activated, causing the CSMA-inspired activation pattern to resemble a fixed reuse pattern.

%However, as $L$ increases, the rate of increase in the geometric means slows. This is because $K_{avg}$ is fixed, so the multicast opportunities are limited even if $L$ becomes very large.

In Figure~\ref{fig:hf_changeL} we plot the goodput CDF for HF for the optimum solution for $L \in \{1,10,20,35,40 \}$. We can draw from Figure~\ref{fig:hf_changeL} a similar conclusion as we did from Figure~\ref{fig:G_changeL}: The performance of the optimum solution, i.e., the minimum rate of the users, increases as $L$ increases, and the rate of this increase slows down. 

For a typical video-coding rate of $2$ Mb/s, a user must receive the video chunks at least $2$ Mb/s to be able to stream. In Figure~\ref{fig:hf_changeL}, the minimum goodput of the users for $L=1$ is $\bar{r}_{min}^{1}\approx0.027$ and for $L=30$ $\bar{r}_{min}^{30}\approx0.04$. So if CC is not employed, each AP must have at least $74$ Mb/s downlink multicast capacity so that each user can stream, whereas if CC is employed with $L=30$, only $ 50$ Mb/s suffices.

\input{Figs/New_pfvshf}

In Figure~\ref{fig:hf_pf} we plot the goodput CDF of both PF and HF for the optimum solution for $L=20$. %, $\gamma=0.1$, $H=10$, and $K_{avg}=200$. 
From Figure~\ref{fig:hf_pf}, we observe that HF improves the minimum goodput of users with respect to PF, at the expense of decreasing the goodput for those users with larger goodputs, as expected. %only $30\%$ of the users have higher rate at the expense
The minimum goodput of the users for HF is $\bar{r}_{min}^{hf}\approx0.038$ and for PF $\bar{r}_{min}^{pf}\approx0.014$. So if HF is employed, each AP must have at least $53$ Mb/s capacity so that each user can stream, whereas $143$ Mb/s capacity is required if PF is employed. Employing PF instead of HF requires a higher AP capacity; however, on the positive side, more than $60\%$ of users can stream at significantly higher quality when adaptive quality coding is used~\cite{lefteris-adaptivequality}.

\begin{figure}[t]
    \centering
    \resizebox{0.73\columnwidth}{!}{%
    
    \begin{tikzpicture}

    \begin{axis}
    [
    % put axis lines at left and bottom
    axis lines = left,
    % control axis labels
    xlabel = \smaller {Number of scheduling slots},
    ylabel = \smaller {Goodput},
    ylabel near ticks,
    xmin=0,
    % control legend position
    legend pos = north east,
    % control size of tick marks (10,20,30,etc)
    ticklabel style={font=\tiny},
    % control major grids
    grid=both,
    major grid style={line width=.2pt,draw=gray!30},
   % xtick={0,50,100,150,200,250,300,350,400,450,500,550,600,650,700,750,800,850,900,950,1000},
   % ytick={0,0.1,0.2,0.3,0.4,0.5,0.6,0.7,0.8,0.9,1,1.1,1.2,1.3,1.4,1.5},
    % control minor grids
    xtick={0,400,601,1000},
    %minor grid style={line width=.1pt, draw=gray!10},
    %minor tick num=5,
    ]
    
    % \addplot[black]
    % table[y=Opt-L1-Y,x=Opt-L1-X]{Figs/CDF_data_L.tex};
    % \addlegendentry{\smaller $L=1$, Analytical}
    % \addplot[black,dashed]
    % table[y=Super-L1-Y,x=Super-L1-X]{Figs/CDF_data_L.tex};
    % \addlegendentry{\smaller $L=1$, Super-user}
    % \addplot[black!50]
    % table[y=Heur-L1-Y,x=Heur-L1-X]{Figs/CDF_data_L.tex};
    % \addlegendentry{\smaller $L=1$, RGA}
    
    % \addplot[green]
    % table[y=Opt-L5-Y,x=Opt-L5-X]{Figs/CDF_data_L.tex};
    % \addlegendentry{\smaller $L=5$, Analytical}
    % \addplot[green,dashed]
    % table[y=Super-L5-Y,x=Super-L5-X]{Figs/CDF_data_L.tex};
    % \addlegendentry{\smaller $L=5$, Super-user}
    % \addplot[green!50]
    % table[y=Heur-L5-Y,x=Heur-L5-X]{Figs/CDF_data_L.tex};
    % \addlegendentry{\smaller $L=5$, RGA}
    
    \addplot[black]
    table[y=U1,x=Slots]{Figs/New_dynamic_TWC.tex};
    \addlegendentry{\smaller $u_1$}
     \addplot[blue]
    table[y=U2,x=Slots]{Figs/New_dynamic_TWC.tex};
    \addlegendentry{\smaller $u_2$}
     \addplot[gray]
    table[y=U3,x=Slots]{Figs/New_dynamic_TWC.tex};
    \addlegendentry{\smaller $u_3$}
     \addplot[red]
    table[y=U4,x=Slots]{Figs/New_dynamic_TWC.tex};
    \addlegendentry{\smaller $u_4$}
     \addplot[orange]
    table[y=U5,x=Slots]{Figs/New_dynamic_TWC.tex};
    \addlegendentry{\smaller $u_5$}
    \addplot[green,dashed]
    table[y=U6,x=Slots]{Figs/New_dynamic_TWC.tex};
    \addlegendentry{\smaller $u_6$}
    \addplot[purple,dashed]
    table[y=U7,x=Slots]{Figs/New_dynamic_TWC.tex};
    \addlegendentry{\smaller $u_7$}
    
%    \addplot[red]
%    table[y=CDF-Reduced,x=Rate-Reduced]{Figs/New_reducedvectors.tex};
%    \addlegendentry{\smaller Reduced vectors}
%    \addplot[black,dashed]
%    table[y=CDF-Approx,x=Rate-Approx]{Figs/New_reducedapprox.tex};
%    \addlegendentry{\smaller Dynamic}

    \end{axis}

    \end{tikzpicture}
    }
    \caption{Example of goodput versus number of scheduling slots for a dynamic network for PF. }
    \label{fig:dynamic}
\end{figure}
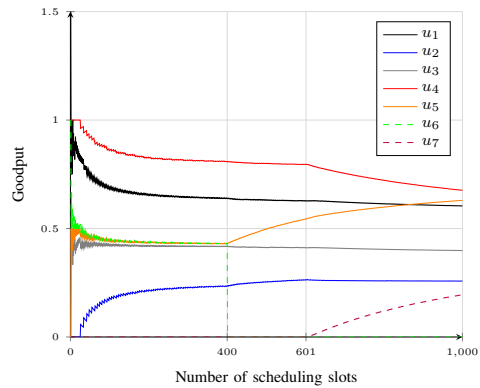

Finally, in Figure~\ref{fig:dynamic}, we plot the goodput versus the number of scheduling slots for a dynamic network for PF. We use the same network setup in Fig.~\ref{fig:network_model}, However, we assume that at scheduling slot $400$, user $u_6$ leaves the network, and at slot $601$, user $u_7$, assigned to $\Phi_1$, joins the network in the intersection of the transmission radii of both APs.
The DPP algorithm~\eqref{queue_eq}-\eqref{instantaneous-wsrm} is applied across the dynamic network transition, and the corresponding user goodputs are calculated as time averages of the instantaneous rates. 
%user goodputs are calculated by solving~\eqref{queue_eq}-\eqref{instantaneous-wsrm}. 
Figure~\ref{fig:dynamic} shows two things: 1) Since the network is the same as the network in Fig.~\ref{fig:network_model} initially, the goodputs of the users converge to their respective optimum goodputs where $\bar{\Br}_{pf}^\star~\approx~(0.62,0.25,0.42,0.83,0.42,0.42)$ (see Table~\ref{tab:equivusertable}) until $400$th slot, when user $u_6$ leaves the network, i.e., when the network topology changes. This confirms that 
applying the DPP algorithm in stationary conditions yields a goodput vector very close to the optimal stationary solution~\eqref{eq:optimization_problem_main}, where ``close’’ is in the sense of the bound in~\eqref{eq:dpp-convergence}. 
%solving \eqref{virtual_arrival}-\eqref{instantaneous-wsrm} at every scheduling slot and updating the user queues by~\eqref{queue_eq} gives the same goodput vector as solving~\eqref{eq:optimization_problem_main}, as expected. 
2) At both $400$th and $601$st slots, user goodputs adapt to the changes in the network and converge to their new optimum goodputs. In particular, the goodput of user $u_6$ goes to $0$ at the $400$th slot since it leaves the network, and the goodput of user $u_7$ becomes nonzero after the $601$st slot since it joins the network.

\section{Conclusion}

Considering coded caching techniques for on-demand video streaming over WLANs where multiple users are served simultaneously by multiple spatially distributed APs, we formulated the region of achievable goodput (defined as %the number of video chunks made available to the user per unit time) 
the number of video chunks per unit time delivered in
the users’ playback buffer) 
and studied the per-user goodput distribution under proportional and hard fairness scheduling. %We also developed reduced complexity and online scheduling strategies that are adaptive to dynamic network conditions and 
We proposed a dynamic scheduling algorithm that provably achieves the optimal fairness point in stationary conditions with reduced complexity, and introduced a heuristic to further reduce the complexity, achieving a favorable
trade-off between performance and complexity. We compared the proposed schemes with standard state-of-the-art techniques, including conventional (uncoded) caching, collision avoidance by allocating APs to different sub-channels (i.e., spatial reuse), and a CSMA-inspired method that assigns timer values to APs from an exponential distribution to determine which APs are activated. Simulation results confirmed the performance of the proposed schemes and showed that employing coded caching for on-demand video streaming gives significant gains.

\bibliographystyle{IEEEtran}
\bibliography{references,kagan_references}

%\clearpage
%\newpage
%\appendix
%\input{Appendix.tex}

\end{document}